\documentclass{aa}
\titlerunning{Carbon cycling and interior evolution of water-covered planets}

\usepackage{graphicx}

\usepackage{dblfloatfix}
\usepackage{floatrow}
\usepackage{lineno}

\usepackage{apalike}

\usepackage{natbib}

\usepackage{savesym}
\usepackage{amsmath}
\savesymbol{iint}
\usepackage{txfonts}
\restoresymbol{TXF}{iint}

\usepackage{wasysym}

\begin{document} 

   \title{Carbon cycling and interior evolution of water-covered plate tectonics and stagnant lid planets}

   \author{Dennis H\"oning
          \inst{1,2}\thanks{Corresponding author}
          \and
         Nicola Tosi
         \inst{3,4}
          \and
          Tilman Spohn
          \inst{3}
          }
   \institute{Origins Center, Nijenborgh 7, 9747 AG Groningen, The Netherlands\\
   \and
             Department of Earth- and Life Sciences, Vrije Universiteit Amsterdam, Amsterdam, The Netherlands\\
             \email{d.hoening@vu.nl}\\
             \and
          Institute of Planetary Research, German Aerospace Center (DLR), Berlin, Germany\\
          \and
          Department of Astronomy and Astrophysics, Technische Universit\"at Berlin, Berlin, Germany
             }


 
  \abstract
   {}
   {The long-term carbon cycle for planets with a surface entirely covered by oceans works differently from that of the present-day Earth because inefficient erosion leads to a strong dependence of the weathering rate on the rate of volcanism. In this paper, we investigate the long-term carbon cycle for these planets throughout their evolution.}
   {We built box models of the long-term carbon cycle based on CO$_2$ degassing, seafloor-weathering, metamorphic decarbonation, and ingassing and coupled them with thermal evolution models of stagnant lid and plate tectonics planets.}
   {The assumed relationship between the seafloor-weathering rate and the atmospheric CO$_2$ or the surface temperature strongly influences the climate evolution for both tectonic regimes. For a plate tectonics planet, the atmospheric CO$_2$ partial pressure is characterized by an equilibrium between ingassing and degassing and depends on the temperature gradient in subduction zones affecting the stability of carbonates. For a stagnant lid planet, partial melting and degassing are always accompanied by decarbonation, such that the combined carbon content of the crust and atmosphere increases with time. Whereas the initial mantle temperature for plate tectonics planets only affects the early evolution, it influences the evolution of the surface temperature of stagnant lid planets for much longer.}
   {For both tectonic regimes, mantle cooling results in a decreasing atmospheric CO$_2$ partial pressure. For a plate tectonics planet this is caused by an increasing fraction of subduction zones that avoid crustal decarbonation, and for stagnant lid planets this is caused by an increasing decarbonation depth. This mechanism may partly compensate for the increase of the surface temperature due to increasing solar luminosity with time and hereby contribute to keep planets habitable in the long-term.}
   

   \maketitle

\section{Introduction}
\label{sec1}
The negative feedback provided by the long-term carbonate-silicate cycle is important in keeping Earth’s climate habitable in the long-term \citep{Walker:1981,Kasting:2003}. Should plate tectonics on planets beyond our solar system work in a similar way as on Earth, these planets could remain habitable over a wide range of orbital distances and long time-spans \citep{Kasting:1993,Franck:2000}. The negative feedback is mainly controlled by temperature-dependent continental weathering, which requires emerged land \citep{Foley:2015}. On the one hand, emerged land allows for a strong temperature dependence of weathering, and on the other hand, it allows for efficient erosion (e.g., \citealt{Flament:2008}), exposing fresh uncarbonated surface for weathering. However, the continental volume on planets beyond our solar system may differ significantly \citep{honing:2019}. Furthermore, in Earth’s early evolution, the continents were largely flooded, resulting in a strongly reduced erosion rate (e.g., \citealt{Flament:2008}), affecting the efficiency of continental weathering.

The fraction of emerged land required to keep the continental weathering feedback efficient may be small, and \citet{Abbot:2012} argue for a strong continental weathering feedback for a continental fraction of more than 1\%. However, without any emerged land, this particular weathering feedback mechanism does not work. A negative feedback mechanism that works for water-covered oceanic crust is provided by seafloor-weathering \citep{Brady:1997,Sleep:2001}. Seafloor-weathering was more important in Earth’s history, when the surface was covered by water, than it is today \citep{Krissansen-Totton:2017}. \citet{Mills:2014} argue for a shift in importance from seafloor to continental weathering 1.5 to 0.5 billion years b.p. and \citet{Coogan:2013} argue that both feedbacks had a similar magnitude in the late Mesozoic. The major difference with respect to continental weathering is that seafloor-weathering does not depend on the erosion rate to provide fresh surface that can be weathered, but rather on the supply rate of fresh basaltic crust by volcanism. On plate tectonics planets, fresh basaltic crust is mainly generated by seafloor spreading. The way seafloor weathering behaves is debated. Seafloor weathering was commonly believed to directly depend on the atmospheric CO$_2$ partial pressure \citep{Brady:1997,Sleep:2001}. However, more recent studies argue that it mainly depends on the pore-space temperature, which in turn is controlled by the planetary surface temperature \citep{Coogan:2015,Krissansen-Totton:2017}. 

Stagnant lid planets lack mid-ocean ridges and subduction zones. The long-term carbonate-silicate cycle operates differently for these bodies than it does for plate tectonics planets. Volcanic degassing occurs at hot spots, and the accumulation of CO$_2$ in the atmosphere affects the habitability of these planets \citep{Noack:2017,Tosi:2017,Godolt:2019}. \citet{Tosi:2017} showed that the oxidation state of the mantle crucially affects the amount of degassed CO$_2$, and the accumulated atmospheric CO$_2$ affects the outer boundary of the habitable zone. Large amounts of atmospheric CO$_2$ allow liquid water to exist on the planetary surface farther away from the star. The inner boundary of the habitable zone is also affected by the atmospheric CO$_2$ if the amount of degassed H$_2$O is small \citep{Tosi:2017}. However, it remains unclear whether carbon recycling can occur on stagnant lid planets and to what extent this can regulate their atmospheric CO$_2$ partial pressure.

The production of fresh crust by volcanism provides a potential carbon sink, which could establish a negative feedback similar to seafloor-weathering \citep{Foley:2018}. Volcanic eruptions may bury carbonated crust, which then sinks deeper into the mantle. Increasing pressure and temperature with depth affect the stability of carbonates. The release of CO$_2$ back into the atmosphere caused by the increase of temperature with depth is known as metamorphic decarbonation (e.g., \citealt{Bickle:1996}). Accounting for burial and decarbonation of the crust, \citet{Foley:2018} showed that for stagnant lid planets with an Earth-like total CO$_2$ budget, carbon cycling could provide a negative feedback avoiding a supply-limited regime, which would occur through a complete carbonation of the crust. It remains unclear, however, whether this negative feedback is sufficiently strong to regulate the climate. Or does CO$_2$ rather accumulate in the atmosphere with time, with weathering hardly influencing the atmospheric CO$_2$ budget?

On planets without plate tectonics, the production of fresh crust occurs via hot-spot volcanism. In the presence of water, the basaltic crust can be carbonated and subsequently buried by new volcanic eruptions. This also implies that even if subsequent erosion of the upper crust took place, it would not expose uncarbonated crust for weathering, in contrast to the present-day Earth. Instead, the crust exposed by erosion was already carbonated at the time it was formed. Therefore, the CO$_2$ sink on a stagnant lid planet is directly coupled to the production rate of new crust and hence to the CO$_2$ source and not to the erosion rate. \citet{Valencia:2018} studied the seafloor-weathering feedback for tidally heated planets. They calculated equilibrium states of the atmosphere and mantle and concluded that the negative feedback may keep these planets habitable if fresh rock is supplied at a sufficiently large rate and seafloor-weathering is temperature-dependent. However, the mantle temperature of a planet that is not tidally heated substantially evolves with time. This is particularly important for the stability of carbonates, which can enter the mantle in substantial quantities only if the temperature-depth gradient is shallow \citep{Foley:2018}. \citet{foley:2019} explored the timespan a stagnant-lid planet can sustain active volcanism and therefore a temperate climate, depending on the planetary CO$_2$ budget and on the internal heating rate. In particular, he showed that the minimum CO$_2$ budget of a planet required to recover from a snowball state crucially depends on whether or not exchange of CO$_2$ between the ocean and the atmosphere is possible.

In this paper, we will focus on the interplay between the mantle temperature and the atmospheric CO$_2$ during planetary evolution and explore factors controlling the accumulation of CO$_2$ in the atmosphere or climate regulation via seafloor-weathering. In particular, we will study the influence of different relationships for seafloor-weathering: Models with seafloor-weathering directly dependent on the CO$_2$ partial pressure \citep{Sleep:2001} and models with seafloor-weathering dependent on the pore-space temperature, which is a function of the surface temperature \citep{Krissansen-Totton:2017}. We will furthermore explore the effects of the tectonic regime (plate tectonics vs. stagnant lid) and other planet specific parameters such as the initial mantle temperature, mantle rheology, and oxidation state of the mantle.

For a plate tectonic planet, carbon recycling into the mantle occurs at subduction zones. The temperature profile of a subduction zone depends on several parameters such as subduction angle and plate speed and varies from one subduction zone to another \citep{Penniston-Dorland:2015}. In hot subduction zones, the temperature can exceed the decarbonation temperature, causing the release of CO$_2$ \citep{Johnston:2011}. Most of this CO$_2$ then finds its way through faults and volcanic units back to the surface and is not recycled into the mantle. In the present-day Earth, most subduction zones are sufficiently cold to avoid decarbonation, which enables carbon recycling into the mantle. Note that \citet{ague:2014} point out that the release of fluids in subduction zones can still drive decarbonation for present-day subduction zones, which we neglect in this paper, for simplicity. In the early Earth, most subduction zones were hot, decarbonation was efficient, and recycling of carbon into the mantle was difficult \citep{Dasgupta:2004,Dasgupta:2010}.

For a stagnant lid planet, carbon recycling into the mantle could potentially occur through continuous volcanism, burial and sinking of carbonated crust, eventually followed by delamination. However, the crustal temperature increases with depth and the slow sinking of carbonated crust causes it to be close to thermal equilibrium with its surrounding. Carbonates are stable up to a certain pressure-dependent temperature above which metamorphic decarbonation releases CO$_2$ that may then find its way back into the atmosphere \citep{Foley:2018}. Whether or not decarbonation occurs, therefore, depends on the temperature profile in the carbonated crust. However, even if decarbonation occurs before the crust reaches the mantle, the slowly sinking crust can store carbon temporarily. After the planet has cooled sufficiently for mantle melting to cease, volcanism and mantle degassing will cease, too. No fresh crustal rock can be produced and weathering will greatly slow down. The atmospheric CO$_2$ partial pressure is then no longer controlled by interior-surface processes and will tend to remain constant. In the following, we derive a model of the atmospheric CO$_2$ partial pressure depending on the tectonic regime and thermal state of the planet.

\section{Carbon cycling and decarbonation on plate tectonic planets}
\label{sec3}

For simplicity, we restrict our analysis to Earth-sized planets. We only consider planets with an entirely water-covered surface such that continental erosion can be neglected. We neglect the ocean as a carbon reservoir (as in \citealt{Foley:2018}), assuming that the climate is regulated by the long-term carbon cycle, however. For water-rich planets with an ocean thickness of several hundred km, the climate will rather be controlled by partitioning of carbon between the atmosphere and the ocean \citep{kite:2018}. Furthermore, we neglect atmospheric escape since Earth-sized planets are not expected to undergo Jeans escape for heavy molecules such as CO$_2$ \citep{CatlingandZahnle:2009}.

The weathering rate for water-covered planets $F_w$ depends on the rate at which fresh crustal volume is produced, $\frac{\mathrm{d}V_{cr}}{\mathrm{d}t}$ and, depending on the scaling used for seafloor-weathering, either on the atmospheric CO$_2$ partial pressure $P_{CO_2}$ or the pore-space temperature $T_p$. We test both models and scale the weathering rate using the present-day Earth seafloor weathering rate $F_{sfw,E}$. For the CO$_2$-dependent weathering rate $F_w^{(CO_2)}$, we have
\begin{linenomath}
\begin{equation}
    F_w^{(CO_2)}=F_{sfw,E}\frac{\mathrm{d}V_{cr}^*}{\mathrm{d}t}\left(P_{CO_2}^*\right)^\alpha, 	
\label{eq:weather1}
\end{equation}
\end{linenomath}
where $\alpha \approx 0.23$ is a constant \citep{Brady:1997,Foley:2015}. The index $E$ denotes present-day Earth variables; variables scaled with their present-day Earth values are noted with an asterisk, i.e. $\frac{\mathrm{d}V_{cr}^*}{\mathrm{d}t}=\frac{\frac{\mathrm{d}V_{cr}}{\mathrm{d}t}}{\frac{\mathrm{d}V_{cr,E}}{\mathrm{d}t}}$ and $P_{CO_2}^*=\frac{P_{CO_2 }}{P_{CO_2,E}}$. The present-day Earth seafloor weathering rate $F_{sfw,E}$ is determined by the present-day Earth ingassing rate $F_{ingas,E}$, assuming that only a fraction $\xi_E$ of the total weathering rate is due to seafloor weathering, and that only fractions $f_E$ and $\phi_E$ will not be removed by arc volcanism and decarbonation, respectively, during subduction:
\begin{linenomath}
\begin{equation}
    F_{sfw,E}=F_{ingas,E}\frac{\xi_E}{f_E \phi_E} 	
\label{eq:sfwE}
\end{equation}
\end{linenomath}
The total flux of carbon into subduction zones $F_{subd}$ linearely depends on the crustal production rate and on the crustal carbon reservoir relative to the present-day Earth value, $R_{crust}^*$, and can be scaled using the present-day Earth seafloor-weathering rate:
\begin{linenomath}
\begin{equation}
    F_{subd}=F_{sfw,E} \frac{\mathrm{d}V_{cr}^*}{\mathrm{d}t} R_{crust}^* 	
\label{eq:subd}
\end{equation}
\end{linenomath}
Note that Eq. \ref{eq:subd} does not include the parameter $\xi_E$, since it describes the carbon flux into subduction zones of a water-covered planet for which seafloor-weathering is the only weathering mechanism. The ingassing rate depends on the total flux of carbon into subduction zones times the fractions that are not removed by arc volcanism ($f$) or decarbonation ($\phi$):
\begin{linenomath}
\begin{equation}
    F_{ingas}=F_{subd} f \phi	
\label{eq:ingas}
\end{equation}
\end{linenomath}
We assume that the present-day Earth degassing and ingassing rates are in equilibrium and scale the degassing rate with the crustal production rate and with the mantle carbon reservoir. Neglecting minor amounts of carbon in the atmosphere and oceans, the mantle carbon reservoir $R_{mantle}$ can be described as the difference between the total carbon reservoir $R_{tot}$ and the crustal carbon reservoir. We have
\begin{linenomath}
\begin{equation}
    F_{degas}=F_{ingas,E}\frac{\mathrm{d}V_{cr}^*}{\mathrm{d}t}R_{mantle}^*,
\label{eq:degas}
\end{equation}
\end{linenomath}
with $R_{mantle}=R_{tot}-R_{crust}$. The rate of change of the crustal carbon reservoir $\frac{d}{dt}R_{crust}$ can be calculated as the difference between the weathering rate (Eq. \ref{eq:weather1}) and the rate at which carbon enters subduction zones (Eq. \ref{eq:subd}):
\begin{linenomath}
\begin{equation}
    \frac{\mathrm{d}}{\mathrm{d}t}R_{crust}=F_{sfw,E}\frac{\mathrm{d}V_{cr}^*}{\mathrm{d}t}\left(\left(P_{CO_2}^*\right)^\alpha-R_{crust}^*\right)
\label{eq:crust}
\end{equation}
\end{linenomath}
The rate of change of the atmospheric carbon is given by $\frac{d}{dt}R_{atm}=F_{subd}-F_{ingas}+F_{degas}-F_{sfw}$, i.e.
\begin{linenomath}
\begin{equation}
\begin{split}
    \frac{\mathrm{d}}{\mathrm{d}t}R_{atm} & =F_{sfw,E}\frac{\mathrm{d}V_{cr}^*}{\mathrm{d}t} \\\ &
    \cdot\left(R_{crust}^*(1-f\phi)+\frac{f_E\phi_E}{\xi_E}(R_{tot}-R_{crust})^*-\left(P_{CO_2}^*\right)^\alpha\right).
\label{eq:atm}
\end{split}
\end{equation}
\end{linenomath}
We can now find steady state values for the crustal and atmospheric reservoirs by setting Eqs. \ref{eq:crust} and \ref{eq:atm} equal to zero. We obtain
\begin{linenomath}
\begin{equation}
    R_{crust}^*=\left(P_{CO_2}^*\right)^\alpha
\label{eq:sscrust}
\end{equation}
\end{linenomath}
and
\begin{linenomath}
\begin{equation}
    \left(P_{CO_2}^*\right)^\alpha=\frac{\Gamma R_{tot}^*}{\frac{R_{crust,E}}{R_{tot,E}}(\Gamma-1)+1},
\label{eq:ssatm}
\end{equation}
\end{linenomath}
with 
\begin{linenomath}
\begin{equation}
    \Gamma=(f^*\phi^*\xi_E)^{-1}.
\label{eq:gamma}
\end{equation}
\end{linenomath}
Note that in steady state, the atmospheric carbon concentration does not depend on the crustal production rate. This is because the ingassing rate and the degassing rate both depend on the crustal production rate in the same manner. Eq. \ref{eq:ssatm} may be used to analytically calculate the atmospheric carbon concentration on water-covered planets with plate tectonics with variable total carbon reservoirs and carbon fractions $f$ and $\phi$ that remain stable during subduction. We will first restrict ourselves to an Earth-like total carbon reservoir ($R_{tot}^*=1$) but also present results varying the total carbon reservoir. Keeping the arc volcanism fraction constant ($f^*=1$) throughout our analysis, we calculate $\phi^*$ dependent on the temperature-depth profiles of subduction zones and a pressure-dependent decarbonation temperature. Upon scaling the model, we use a relative present-day Earth crustal carbon reservoir ($\frac{R_{crust,E}}{R_{tot,E}}$) of 10\% that is consistent with \cite{Sleep:2001}.

In order to model a $T_p$-dependent seafloor-weathering rate $F_w^{(T_p)}$, we use an Arrhenius expression following \citet{Krissansen-Totton:2017}
\begin{linenomath}
\begin{equation}
    F_w^{(T_p)}=F_{sfw,E}\frac{\mathrm{d}V_{cr}^*}{\mathrm{d}t} \mathrm{exp} \left(\frac{E_{bas}}{R T_{p,E}}-\frac{E_{bas}}{R T_{p}}\right), 	
\label{eq:weather2}
\end{equation}
\end{linenomath}
where $E_{bas}$ is the activation energy for basalt weathering, combined with an empirically derived linear relationship between the pore-space temperature and the surface temperature $T_p=a_{grad}T_s+b_{int}+9$ with the constants $a_{grad}=1.02$ and $b_{int}=-16.7$ \citep{Krissansen-Totton:2017}. Eqs. \ref{eq:sfwE}-\ref{eq:degas} remain unchanged, while the CO$_2$-dependent term $\left(P_{CO_2}^*\right)^\alpha$ in Eqs. \ref{eq:crust}-\ref{eq:ssatm} is replaced by the temperature-dependent term $\mathrm{exp} \left(\frac{E_{bas}}{R T_{p,E}}-\frac{E_{bas}}{R T_{p}}\right)$.

In both models, the carbon fraction $\phi$ that remains stable during subduction is a major parameter impacting the atmospheric CO$_2$ partial pressure and can be calculated using the decarbonation temperature. \citet{Foley:2018} combined a phase diagram for carbonate-bearing oceanic metabasalt \citep{Kerrick:2001} with the Holland and Powell DS622 thermodynamic database \citep{Holland:2011}. They found that the major part of metamorphic decarbonation occurs in a narrow temperature interval by the breakdown of dolomite. We follow their approach and assume that decarbonation occurs at a temperature $T_{decarb}$ increasing linearly with depth and given by
\begin{linenomath}
\begin{equation}
         T_{decarb}=Az+B,
    \label{eq:decarb}
\end{equation}
\end{linenomath}
where $z$ is the depth in m, $A$=3.125$\cdot 10^{-3}$  K/m, and $B$=835.5 K \citep{Foley:2018}.

The fraction of stable carbon $\phi$ depends on how the temperature of the carbonated crust increases on its way into the mantle. However, temperature-depth profiles in subduction zones may differ significantly. While in hot subduction zones crustal decarbonation may occur, relatively cold subduction zones allow carbonated crust to reach the mantle. The global carbon fraction $\phi$ that remains stable during subduction depends on the distribution of the temperature-depth profiles of the actual subduction zones. Since the fraction $\phi$ may change with time, we derive a parameterization dependent on the thermal state of the planet.

The temperature gradient in subduction zones depends on various parameters, such as the mantle temperature, plate speed, subduction angle, and others. For most subduction zone parameters, it is unclear how they scale with planet evolution. For example, the plate speed increases with the convection strength as derived from boundary layer theory (e.g., \citealt{Schubert:2001}), which would yield a larger plate speed when the Earth was hotter, but at the same time it decreases with the crustal thickness, which has also been larger in the past \citep{Sleep:1982}, possibly causing sluggish plate tectonics \citep{Korenaga:2006}. Altogether, the exact dependence of the plate speed on the thermal state of the planet is difficult to predict. In contrast, it is clear that the temperature gradient of a specific subduction zone directly depends on the mantle temperature \citep{England:2004}. Therefore, we set the temperature increase along the subducting slab from the surface to a specific depth $z'$, $T_{subd,z'}-T_s$, to be proportional to the average temperature increase between the surface and the mantle, $T_m-T_s$:
\begin{linenomath}
\begin{equation}
     T_{subd,z'}-T_s=c(T_m-T_s)     
    \label{eq:subdtemp}
\end{equation}
\end{linenomath}
The constant $c$ represents the specific subduction zone (i.e., hot or cold). We keep the depth $z'$ constant at 80 km, which is approximately the depth up to which a strong temperature gradient for the top of a subducting slab can be maintained (c.f. \citealt{Penniston-Dorland:2015} and references therein). We take a present-day distribution of subduction zone temperature gradients between 5 K/km and 10 K/km. The temperature gradients in some subduction zones may be larger \citep{Penniston-Dorland:2015}. However, such temperature gradients cannot be maintained to depths of 80 km. The temperature increase of the hottest subduction zones in \citet{Penniston-Dorland:2015} is not larger than 800 K, which corresponds to an average temperature gradient of 10 K/km throughout the upper 80 km. Using a present-day surface and upper mantle temperature of 285 K and 1650 K, respectively,  we obtain $c$ values between 0.3 and 0.6. A small value of $c$ represents a rather cold subduction zone compared to the others while a large value represents a hot subduction zone. We assume that $c$ is uniform and constant in time, so the temperature of each subduction zone increases linearly with the mantle temperature. Depending on the thermal state of the planet, we can derive a critical value $c_{crit} (t)$ that determines whether or not a particular subduction zone allows subducted carbonates to reach the mantle.

\begin{figure}
\centering
\includegraphics[width=9cm]{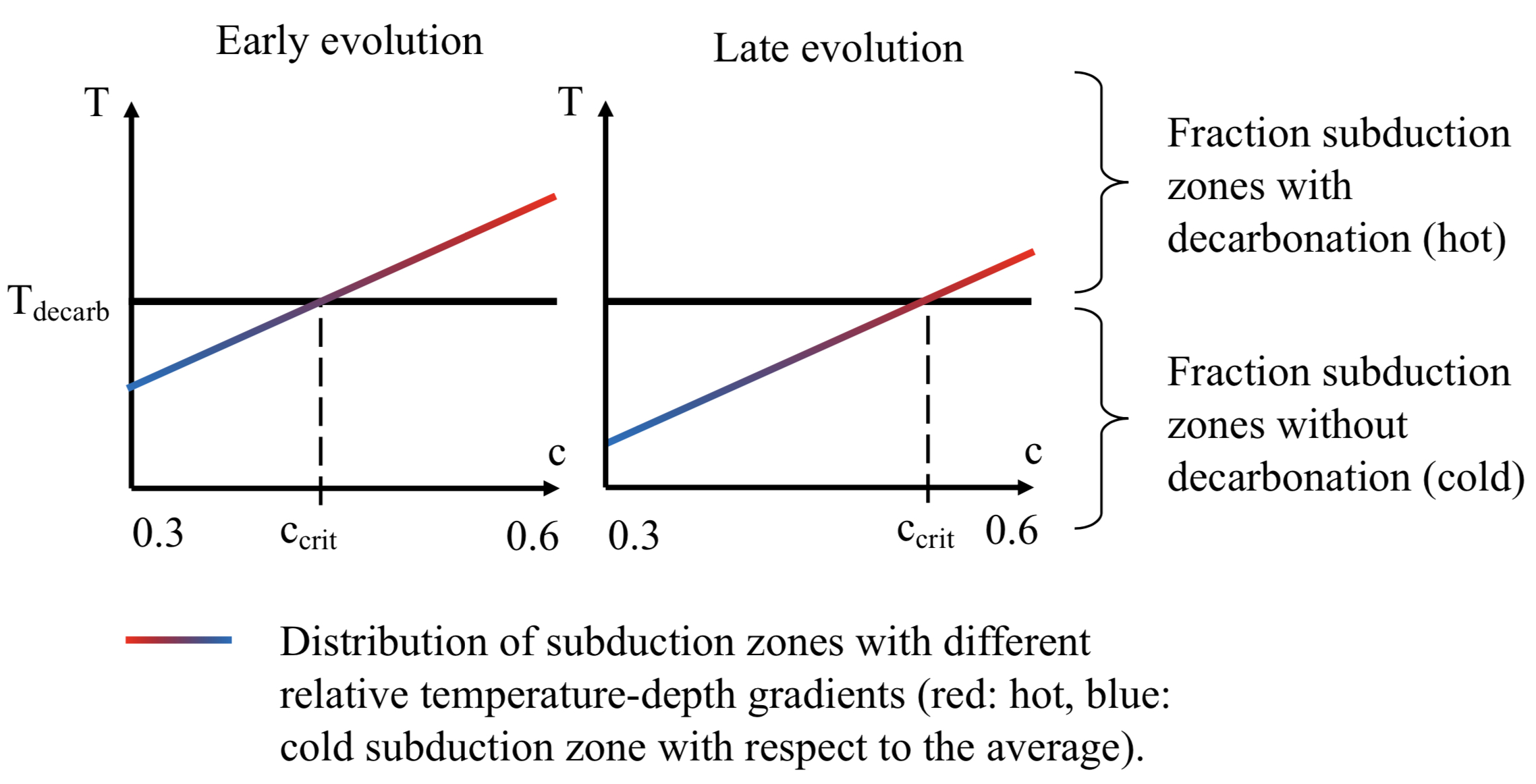}
\caption{Schematic cartoon depicting the distribution of the temperature in the upper mantle of different subduction zones with respect to the decarbonation temperature in the early and late evolution.}
\label{fig1}
\end{figure}

The fraction of $c<c_{crit} (t)$ combines subduction zones that are sufficiently cold to avoid decarbonation while the fraction of $c>c_{crit} (t)$ combines hot subduction zones in which decarbonation takes place (see Fig. \ref{fig1}). Note that $c_{crit}$ increases as the mantle cools, such that an increasing fraction of subduction zones avoid decarbonation.
    
Since the temperature-depth gradient for the top of a subducting slab is particularly large within the first 80 km, we assume that if decarbonation takes place, this occurs before the carbonated crust reaches this depth. If the temperature of a specific subduction zone at 80 km exceeds the decarbonation temperature, decarbonation occurs. On the contrary, if the temperature at 80 km does not exceed the decarbonation temperature, carbon can be recycled into the mantle. The threshold value $c_{crit}$, which determines whether or not decarbonation occurs in a specific subduction zone, can thus be obtained by setting the temperature of the top of a subducting slab at 80 km $T_{subd,z'}$ equal to the decarbonation temperature at this depth. Combining Eqs. \ref{eq:decarb} and \ref{eq:subdtemp}, we have
\begin{linenomath}
\begin{equation}
     c_{crit}(T_m)=\frac{Az+B-T_s}{T_m-T_s},    
    \label{eq:ccrit}
\end{equation}
\end{linenomath}
where the depth $z=z'=80$ km.

The evolution of the mantle temperature is calculated using a parameterized model of mantle convection. Details of this model can be found in Appendix \ref{a1}. We can now calculate the threshold value $c_{crit}(T_m)$, which determines whether or not decarbonation occurs in a specific subduction zone. To obtain the global fraction $\phi$ of carbon that remains stable during subduction, we calculate the fraction of subduction zones for which $c<c_{crit}$. Assuming, for simplicity, a uniform distribution of $c$ between $0.3$ and $0.6$, we can calculate $\phi$ dependent on $c_{crit}$ as
\begin{linenomath}
\begin{equation}
     \phi=\frac{c_{crit}-0.3}{0.6-0.3},   
    \label{eq:phi}
\end{equation}
\end{linenomath}
with $0.3 \leq c_{crit}\leq 0.6$. In order to remain consistent with our scaling to the present-day Earth, we also follow  Eqs. \ref{eq:ccrit} and \ref{eq:phi} when calculating the present-day value $\phi_E$ of the stable global carbon fraction.

In Section \ref{sec51}, we first show examples obtained using fixed surface temperatures of 300 K and 400 K, respectively. However, it is well known that CO$_2$ is a greenhouse gas, which exerts a feedback on the surface temperature and on the CO$_2$ partial pressure. To close the feedback cycle, we use a simple parameterized climate model based on \citet{Walker:1981} (see Appendix \ref{a3}).

\section{Carbon cycling and decarbonation on stagnant lid planets}
\label{sec4}
On plate tectonics planets, CO$_2$ degassing from the mantle mainly occurs via melting at mid-ocean ridges and is therefore independent of the temperature gradient affecting carbon subduction, which is a reason why carbon can be recycled into the mantle despite volcanic activity. In contrast, degassing on stagnant lid planets occurs at hot spots. Partial melting, which ultimately causes degassing, occurs at the transition from the lid to the upper mantle. On its way into the mantle, the buried carbonated crust will pass through the zone at which partial melting takes place. Comparing the decarbonation temperature with the solidus temperature of dry peridotite (e.g., \citealt{Takahashi:1990}) reveals that for depths of up to 250 km, the dry solidus temperature exceeds the decarbonation temperature by more than 500 K. As pointed out by \citet{Foley:2018}, the main fraction of carbonates becomes unstable throughout this narrow temperature interval, and only a minor part of carbonates is stable at significantly larger temperatures. Although plumes may increase the temperature beneath the stagnant lid locally by the temperature difference across the bottom boundary layer, this increase is much smaller. The initial temperature drop of 200 K assumed in \citet{Tosi:2017} decreases rapidly with ongoing evolution. Sinking carbonate in cold downwellings will have a lower temperature than the laterally average temperature, but the temperature difference should also be in the range, however. The presence of water reduces the solidus temperature. However, large amounts of water would be required to substantially reduce the solidus. A solidus reduction of 100 K would already require a bulk water concentration of more than 200 ppm \citep{Katz:2003}. Maintaining sufficient amounts of water in the long-term without water recycling to significantly decrease the solidus would require a large initial mantle water concentration. 

Altogether, stagnant lid planets that are sufficiently hot for partial melting to take place should be inevitably too hot to allow carbon entering the mantle in substantial quantities. In other words, as long as the mantle temperature of a stagnant lid planet is so large that partial melting takes place regionally, metamorphic decarbonation will occur globally. On the other hand, if the mantle temperature is too low for partial melting to occur and if the crust is gravitationally stable, decarbonation will not occur. This is because the crust will only grow in thickness and carbonated crust will only reach the decarbonation depth if fresh crust can be produced by partial melting. It has been argued that gravitational instabilities could recycle the crust into the mantle \citep{ORourke:2012,Johnson:2014}, which could cause sinking and decarbonation of part of the crust for planets in their late evolution even if partial melting does no longer take place. For simplicity, we neglect gravitational instabilities in the carbonated crust in our model.

The rates of change of atmospheric carbon reservoir $R_{atm}$ and crustal carbon reservoir $R_{crust}$ of a stagnant lid planet can be calculated as
\begin{linenomath}
\begin{equation}
     \frac{\mathrm{d}}{\mathrm{d}t}R_{atm}=F_{decarb}-F_{w}+F_{degas}
    \label{eq:atmsl}
\end{equation}
\end{linenomath}
and
\begin{linenomath}
\begin{equation}
     \frac{\mathrm{d}}{\mathrm{d}t}R_{crust}=F_{w}-F_{decarb},
    \label{eq:crustsl}
\end{equation}
\end{linenomath}
where $F_w$, $F_{decarb}$, and $F_{degas}$ are the weathering, decarbonation, and degassing rates, respectively. If partial melting and decarbonation occur in stagnant lid planets, Eqs. \ref{eq:atmsl} and \ref{eq:crustsl} imply that a combined steady state of the atmospheric and crustal carbon reservoirs can only be reached if the mantle degassing rate is zero, which would be the case if all of the carbon is stored in the atmosphere and crust. The reason is that a complete decarbonation implies zero ingassing, and therefore, the steady state would require zero degassing as well. However, it is unlikely that stagnant lid planets can reach this steady state. Rather, the combined degassed CO$_2$ reservoir, which is distributed between the crust and the atmosphere, will increase with time as long as volcanism is active.

\begin{figure}
\centering
\includegraphics[width=9cm]{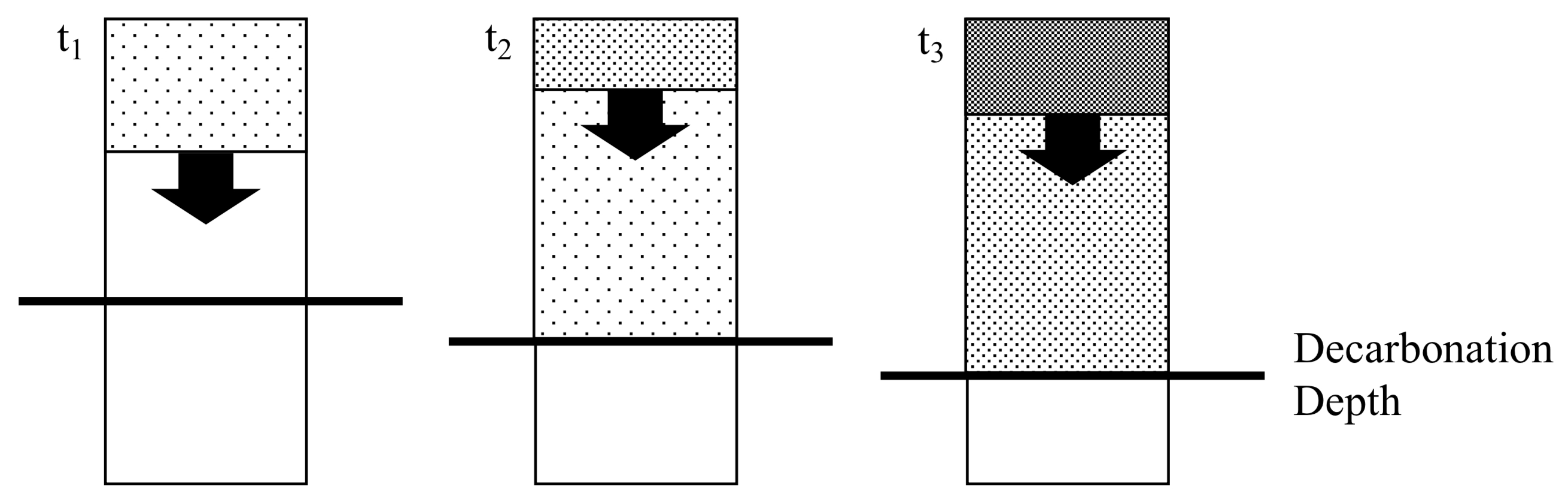}
\caption{Schematic cartoon depicting carbonated crustal burial and sinking for a stagnant lid planet. The dots qualitatively represent the crustal carbon concentration.}
\label{fig2}
\end{figure}

Fig. \ref{fig2} qualitatively illustrates carbonation, burial and sinking of the carbonated crust, and decarbonation. For simplicity, our model assumes that all the carbon is initially stored in the mantle. Therefore, in the early evolution ($t_1$), the CO$_2$ flux to the atmosphere only consists of mantle degassing. The weathering rate, and therefore the crustal carbon concentration, remain small. Later in the evolution ($t_2$), when carbonated crust reaches the decarbonation depth, the CO$_2$ flux to the atmosphere is given by the sum of the degassing rate and the decarbonation rate, the latter in turn is determined by the weathering rate at the time the crust was produced. As a result, the weathering rate and the carbon concentration of freshly produced crust increase. Note that an opposite effect is the increase of the decarbonation depth upon mantle cooling, which causes the decarbonation rate to diminish.

Weathering is a surface-related process and we assume that the growth of the crust is so slow that every finite layer can be carbonated before the next layer is added. Should massive volcanic eruptions produce large volumes of basalt so rapidly that the underlying crust will not be carbonated, we may overestimate the weathering rate and the crustal carbon concentration with this approach. In assuming that the entire formed crust can be subject to weathering, we follow an approach that is typically made for plate tectonics planets \citep{Sleep:2001}.

The evolution of the atmospheric CO$_2$ reservoir is given by Eq. \ref{eq:atmsl}. We follow Eqs. \ref{eq:weather1} and \ref{eq:sfwE} and use $F_{ingas,E}=F_{degas,E}=\frac{\mathrm{d}V_{cr,E}}{\mathrm{d}t}X_E$, where $X_E$ is the present-day Earth mid-ocean ridge CO$_2$ concentration in the melt. For the CO$_2$-dependent weathering scaling we obtain
\begin{linenomath}
\begin{equation}
   F_w^{(CO_2)}=\frac{X_E \xi_E}{f_E \phi_E}\frac{\mathrm{d}V_{cr}}{\mathrm{d}t} \left(P_{CO_2}^*\right)^\alpha.
    \label{eq:weathersl}
\end{equation}
\end{linenomath}
We use $X_E$=125 ppm (e.g., \citealt{Burley:2015}) and a $P_{CO_2,E}=4 \cdot 10^{-4}$ bar. We again test the $T_p$-dependent weathering model by replacing the CO$_2$-dependent term $\left(P_{CO_2}^*\right)^\alpha$ in Eq. \ref{eq:weathersl} with the temperature-dependent term $\mathrm{exp} \left(\frac{E_{bas}}{R T_{p,E}}-\frac{E_{bas}}{R T_{p}}\right)$.

\begin{table*}
\caption[]{Parameter values for the carbon cycle model. $^{(1)}$ from \citet{Foley:2015}, $^{(2)}$ from \citet{Krissansen-Totton:2017}, $^{(3)}$ from \citet{Sleep:2001}, $^{(4)}$ from \citet{Foley:2018}.}
\label{tab1}
\centering                          
\begin{tabular}{l p{9.5cm}l l}        
\hline
Parameter & Description & Value \\
\hline
$A $   & Decarbonation constant       & 3.125 $\cdot 10^{-3}$  K  $^{(4)}$\\
$B $   & Decarbonation constant       & 8.355 $\cdot 10^{2}$  K m $^{-1}$ $^{(4)}$\\

$\alpha$   & Seafloor weathering constant                                   & 0.23  $^{(1)}$\\
$P_{CO_2,E} $ & Present-day Earth atmospheric CO$_2$ partial pressure     & 4 $\cdot 10^{-10}$  bar  \\
$a_{grad} $ & Constant for calculating the pore-space temperature & 1.02 $^{(2)}$  \\ $b_{int} $ & Constant for calculating the pore-space temperature & -16.7 $^{(2)}$  \\
$E_{bas} $ & Activation energy for seafloor weathering & $9.2 \cdot 10^4$ J mol$^{-1}$ $^{(2)}$  \\
$F_{sfw,E}$ & Present-day Earth seafloor weathering rate & $1.75 \cdot 10^{12}$ mol yr$^{-1}$ $^{(2)}$  \\
$\xi_E$ & Seafloor weathering fraction of the total present-day weathering rate  & 0.15  $^{(1)}$ \\
$f_E$ & Present-day Earth fraction of carbon that is not removed by arc volcanism during subduction  & 0.5  \\
$\frac{R_{crust,E}}{R_{tot,E}}$ & Present-day Earth fraction of carbon in the crust  & 0.1  $^{(3)}$ \\
$R_{tot,E}$ & Present-day Earth mantle carbon reservoir & 2.5$\cdot10^{22}$ mol $^{(3)}$ \\
$\omega$ & Relative release of the volatile CO$_2$ reservoir in the crustal matrix to the atmosphere per timestep & 0.02 \\

\hline
\end{tabular}
\end{table*}

For simplicity, we calculate the decarbonation depth assuming a linear temperature profile, thereby neglecting effects of spherical geometry and heat sources in the crust. The temperature throughout the lid with a thickness $D_l$ and boundary layer with a thickness $\delta_m$ is then given by
\begin{linenomath}
\begin{equation}
     T(z)=T_s+\frac{z(T_m-T_s) }{D_l+\delta_m }.
    \label{eq:tzsl}
\end{equation}
\end{linenomath}
Combining Eqs. \ref{eq:decarb} and \ref{eq:tzsl}, the decarbonation depth is
\begin{linenomath}
\begin{equation}
    z_{decarb}=\frac{T_s-B}{A-\frac{T_m-T_s}{D_l+\delta_m }}.
    \label{eq:zdecarbsl}
\end{equation}
\end{linenomath}
To obtain the evolution of the mantle temperature, degassing rate and crustal production rate of a stagnant lid planet we employ a 1D-thermal evolution model based on a parameterization of heat transport by convection (e.g., \citealt{Grott:2011,Tosi:2017}). Details of these models can be found in Appendix \ref{a2} (see also \citealt{Tosi:2017}) and parameter values are given in Tab. \ref{tab1}. The numerical model solves the energy conservation equations of the core, mantle, and stagnant lid. Crustal growth and extraction of incompatible elements occurs via partial melting, and the melt fraction is calculated by comparing the temperature at each depth with the solidus and liquidus temperatures. The mantle viscosity is a function of temperature and water concentration. Since we do not model the water cycle in this paper, we keep the mantle water concentration constant. For the reference case, we use a water concentration of 500 ppm, which is the initial mantle water concentration used in \citet{Tosi:2017}, but also test a smaller water concentration of 125 ppm. We model CO$_2$ degassing based on a model of redox melting following \citet{Grott:2011} calculating the solubility of CO$_2$ in the melt dependent on the oxidation state of the mantle. For the reference case, we use an oxygen fugacity of the iron-w\"ustite buffer (IW), but also test a larger oxygen fugacity of IW+1.

Our numerical model tracks each layer of carbonated crust on its way to decarbonation depth individually and does not assume mixing of carbon in the crust. Whenever a certain layer exceeds decarbonation depth, the carbon content of this layer is set to zero and the CO$_2$ is supplied to the atmospheric reservoir. However, for an increasing decarbonation depth with time, the lowermost carbonated layer can alternately be subject to decarbonation or remain stable. If the atmosphere-crust equilibrium timescale is fast compared to the temporal resolution of the numerical model, this implies an alternating carbon concentration of the newly produced crust. When this crustal layer in turn is subject to decarbonation, the oscillations can amplify and cause numerical instabilities. To avoid these instabilities, we include a delay of the decarbonation flux such that the decarbonated layer releases its CO$_2$ gradually rather than instantanously to the atmosphere. For this purpose, we introduce a temporary volatile CO$_2$ reservoir within crustal matrix, which is fed by the decarbonation flux and which releases at each timestep a fixed proportion $\omega$ of its CO$_2$ reservoir to the atmosphere. Implementing such a delay is reasonable to first order: The CO$_2$ flux throughout the crustal matrix depends on continuous forming and closure of pores and cracks in the crust and on three-dimensional inhomogeneties not captured by our parameterized model. Small values of $\omega$ imply on the one hand efficient damping of oscillations and a smooth atmospheric CO$_2$ curve, but on the other hand a delay in the supply of CO$_2$ to the atmosphere. For our models, we choose a value of $\omega=0.02$, which yields both a smooth curve and a relatively small delay. After $\approx$80 timesteps (or 0.8 Myr using a temporal resolution of 0.01 Myr ), 50\% of the CO$_2$ released by decarbonation is already supplied to the atmosphere. In presenting our results, we will illustrate this delay.

\section{Results}
\label{sec5}

\subsection{Plate tectonics planets}
\label{sec51}

\begin{figure*}
\floatbox[{\capbeside\thisfloatsetup{capbesideposition={right,top},capbesidewidth=5cm}}]{figure}[\FBwidth]
     {\includegraphics[width=12.7cm]{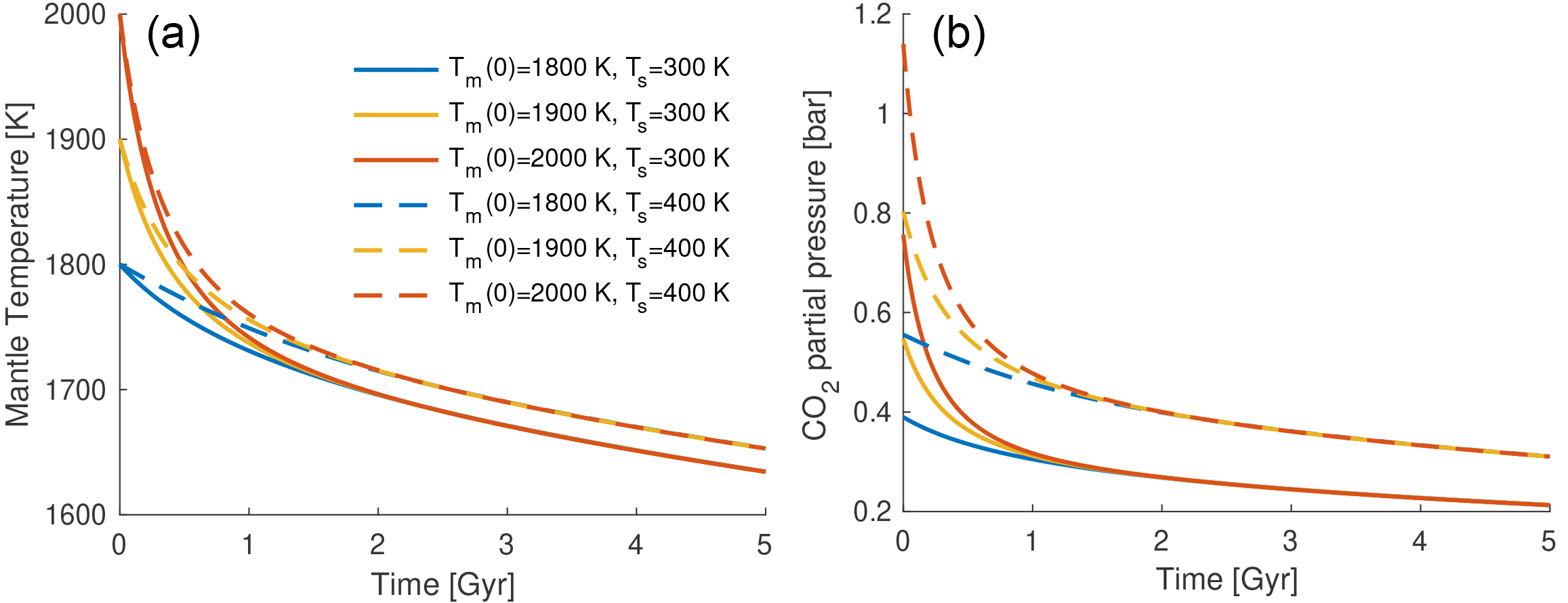}}
   {\caption{Evolutions of (a) the mantle temperature and (b) the atmospheric CO$_2$ partial pressure keeping the surface temperature constant at 300 K (solid) and 400 K (dashed). We use initial mantle temperatures of 2000 K (red), 1900 K (yellow), and 1800 K (blue). We use the CO$_2$-dependent weathering steady-state model with $\alpha=0.23$.}
   \label{fig_constts}}
\end{figure*}

\begin{figure*}[!b]
   \centering
   \includegraphics[width=18cm]{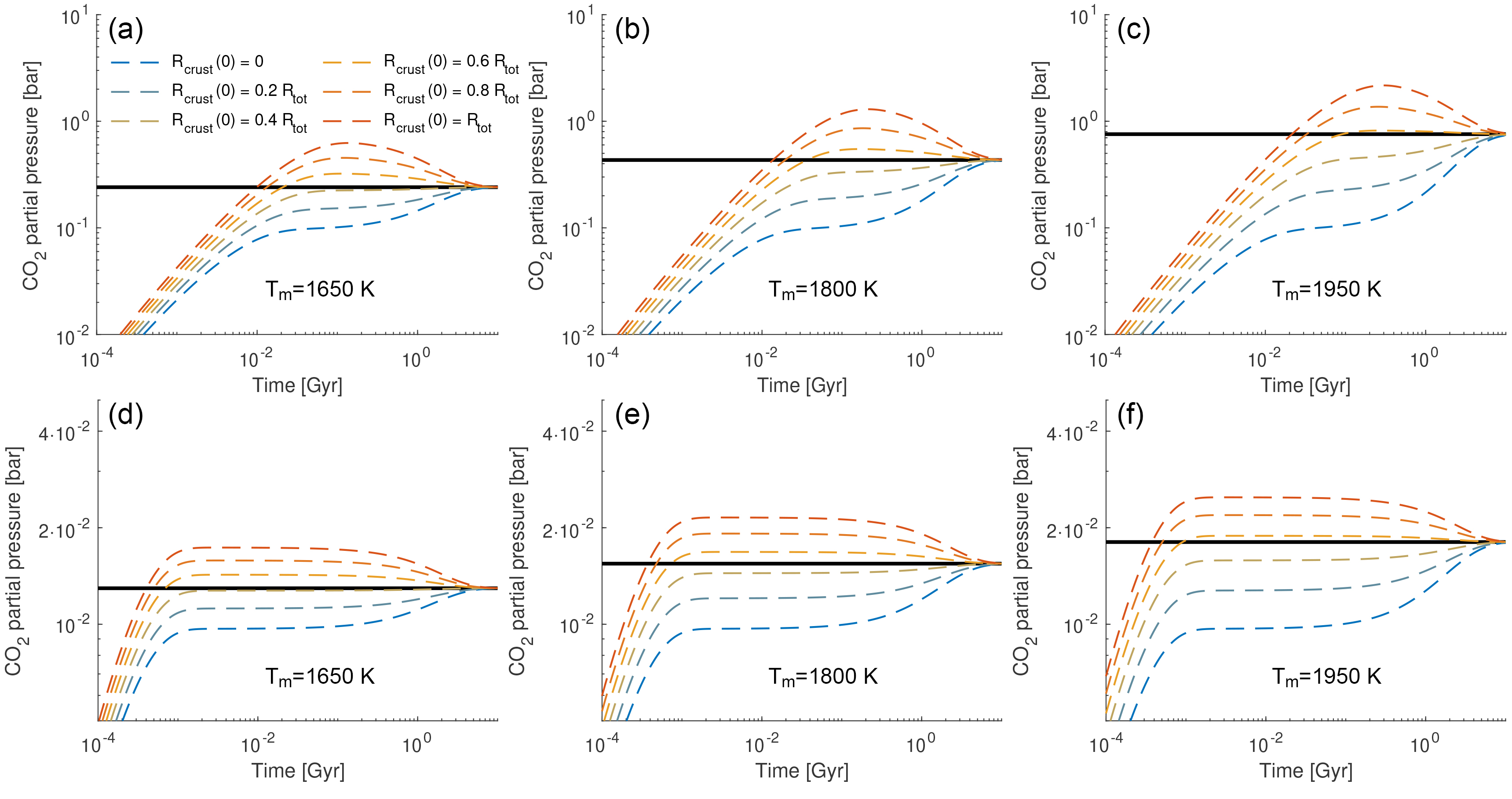}
   \caption{Black solid lines: Equilibrium atmospheric CO$_2$, dashed lines: Evolution curves of the atmospheric CO$_2$ using initial crustal reservoirs of 0, 0.2, 0.4, 0.6, 0.8, and 1 times the total carbon reservoir. The mantle temperature is kept fixed at at 1650 K (a and d), 1800 K (b and e), and 1950 K (c and f), respectively. (a-c) CO$_2$-dependent weathering model using $\alpha=0.23$, (d-f) $T_p$-dependent weathering model.}
   \label{fig_conv}
\end{figure*}

As the mantle cools and an increasing fraction of subduction zones avoid decarbonation, the equilibrium atmospheric CO$_2$ partial pressure decreases. To illustrate this effect, we plot in Fig. \ref{fig_constts} the evolution of (a) the mantle temperature and (b) atmospheric CO$_2$ partial pressure for initial mantle temperatures of 1800 K (blue), 1900 K (yellow), and 2000 K (red) keeping the surface temperature constant at 300 K (solid) and 400 K (dashed). We use a CO$_2$-dependent weathering model with $\alpha$=0.23. Differences due to the different initial mantle temperature vanish at $\approx$1 Gyr. The fixed surface temperature is a boundary condition to the thermal evolution of the mantle and to the decarbonation depths (compare the solid and dashed lines). Accordingly, the atmospheric CO$_2$ approaches different states depending on the surface temperature.

CO$_2$ is a greenhouse gas affecting the surface temperature. The fixed surface temperature in Fig. \ref{fig_constts} does not account for this feedback. In Fig. \ref{fig_conv}, we apply a simple climate model with the surface temperature dependent on the atmospheric CO$_2$ partial pressure for a planet orbiting a Sun-like star (Appendix \ref{a3}). We show the influence of different mantle temperatures, which we keep fixed (a and d: 1650 K, b and e: 1800 K, c and f: 1950 K), on the atmospheric CO$_2$ partial pressure using (a-c) a CO$_2$-dependent weathering model with $\alpha$=0.23 and (d-f) a $T_p$-dependent weathering model. In addition, we show that the atmospheric CO$_2$ obtained from the steady-state models (Eq. \ref{eq:ssatm}), plotted as solid black lines, are stable fixed points towards which the atmospheric CO$_2$ partial pressure evolves. The time-dependent models (dashed lines) start with relative crustal carbon reservoirs of 0, 0.2, 0.4, 0.6, 0.8, and 1 of the total carbon budget (the rest being in the mantle). The initial atmospheric CO$_2$ concentration is zero. However, this only affects the very early evolution ($<$ 0.1 Gyr). The time required to reach a steady state regarding different initial crustal carbon reservoirs is much longer. The models that use the same fixed mantle temperature start to converge at approximately 0.1 Gyr. Neglecting the extreme initial conditions, i.e. all carbon in the mantle or all carbon in the crust (lower and upper curve of each color), the differences of initial conditions after 1 Gyr are small, and differences of initial conditions at the age of the solar system have vanished. Note that the convergence rate is determined by the crustal production rate, which we keep constant. If the carbon cycle operated faster in the early evolution, equilibrium will be reached faster, accordingly.

\begin{figure*}
   \centering
   \includegraphics[width=14.2cm]{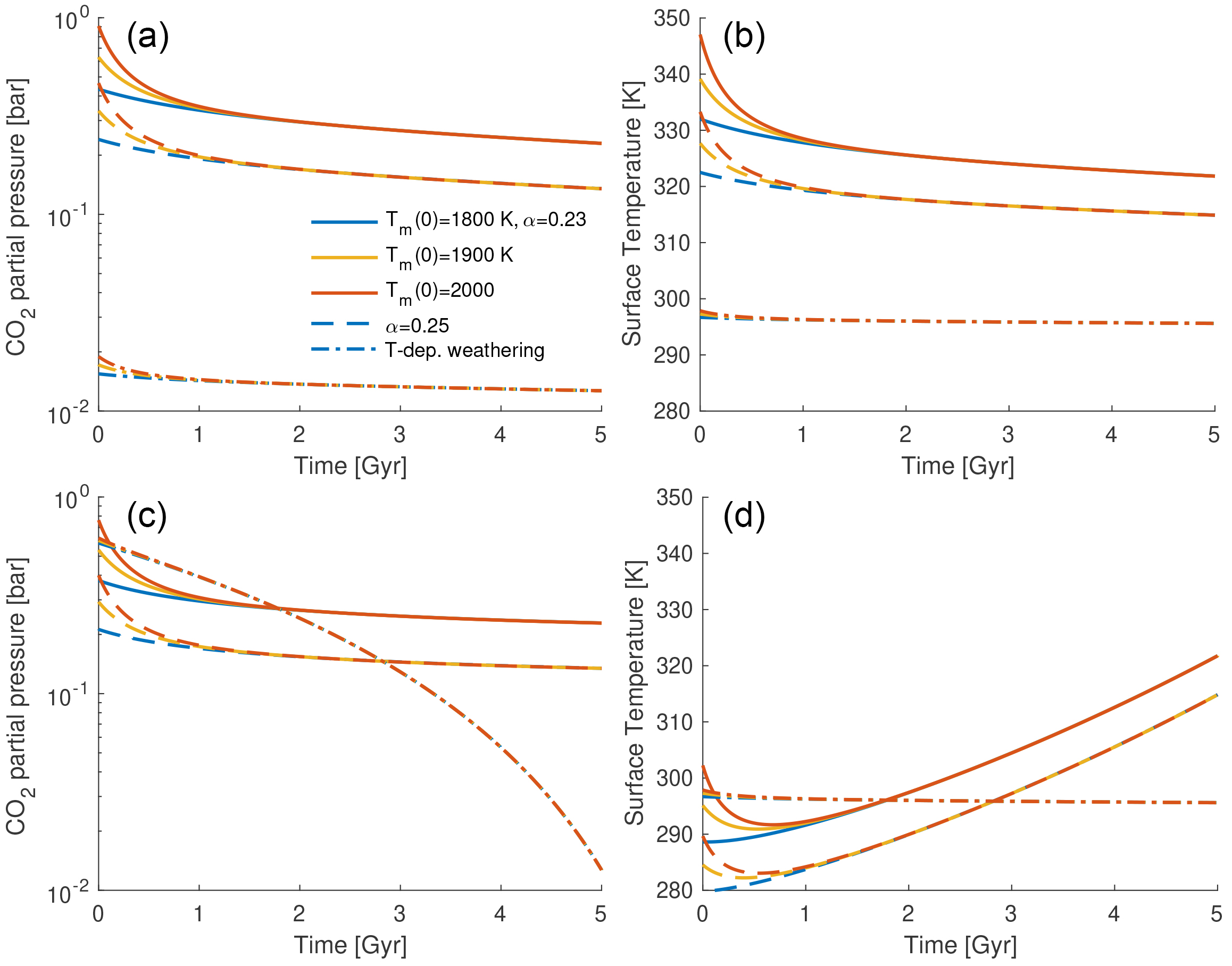}
   \caption{Evolutions of (a and c) atmospheric CO$_2$ and (b+d) surface temperature using initial mantle temperatures of 2000 K (red), 1900 K (yellow), and 1800 K (blue). In (a and b) we keep the solar luminosity constant while in (b and c) the solar luminosity is assumed to increase according to stellar evolution models by a factor of 1.4 in 4.5 Gyr. We use equilibrium models of CO$_2$ depending weathering with $\alpha=0.23$ (solid) and $\alpha=0.25$ (dashed) and equilibrium models of $T_p$-depending weathering (dashed-dotted).}
   \label{fig_ptruns}
\end{figure*}

In Fig. \ref{fig_ptruns}, we use the equilibrium models of CO$_2$-dependent weathering with $\alpha$=0.23 (solid) and $\alpha$=0.25 (dashed) and of $T_p$-dependent weathering (dashed-dotted) and let the mantle and surface temperatures evolve self-consistently. While we keep the solar luminosity constant in Figs. \ref{fig_ptruns}a and \ref{fig_ptruns}b, we use a simple stellar evolution model to calculate increasing solar luminosity (Appendix \ref{a3}) in plotting Figs. \ref{fig_ptruns}c and \ref{fig_ptruns}d. It becomes apparent that the scaling used for seafloor weathering largely determines the evolution of the atmospheric CO$_2$ and thereby the surface temperature.

\begin{figure*}
   \centering
   \includegraphics[width=16cm]{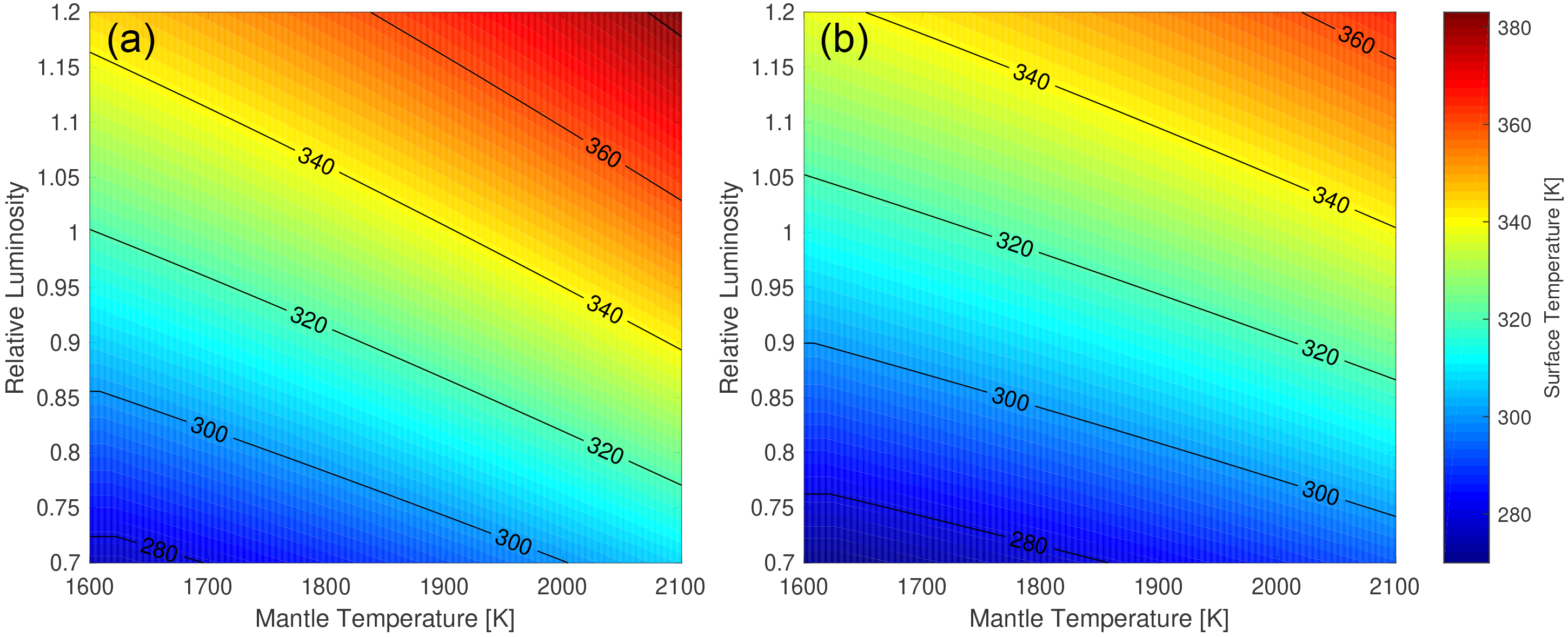}
   \caption{Surface temperature as a function of the relative solar luminosity and mantle temperature for CO$_2$-pressure dependent weathering using (a) $\alpha$=0.23 and (b) $\alpha$=0.25.}
   \label{fig_co2dep}
\end{figure*}

\begin{figure}
   \centering
   \includegraphics[width=9cm]{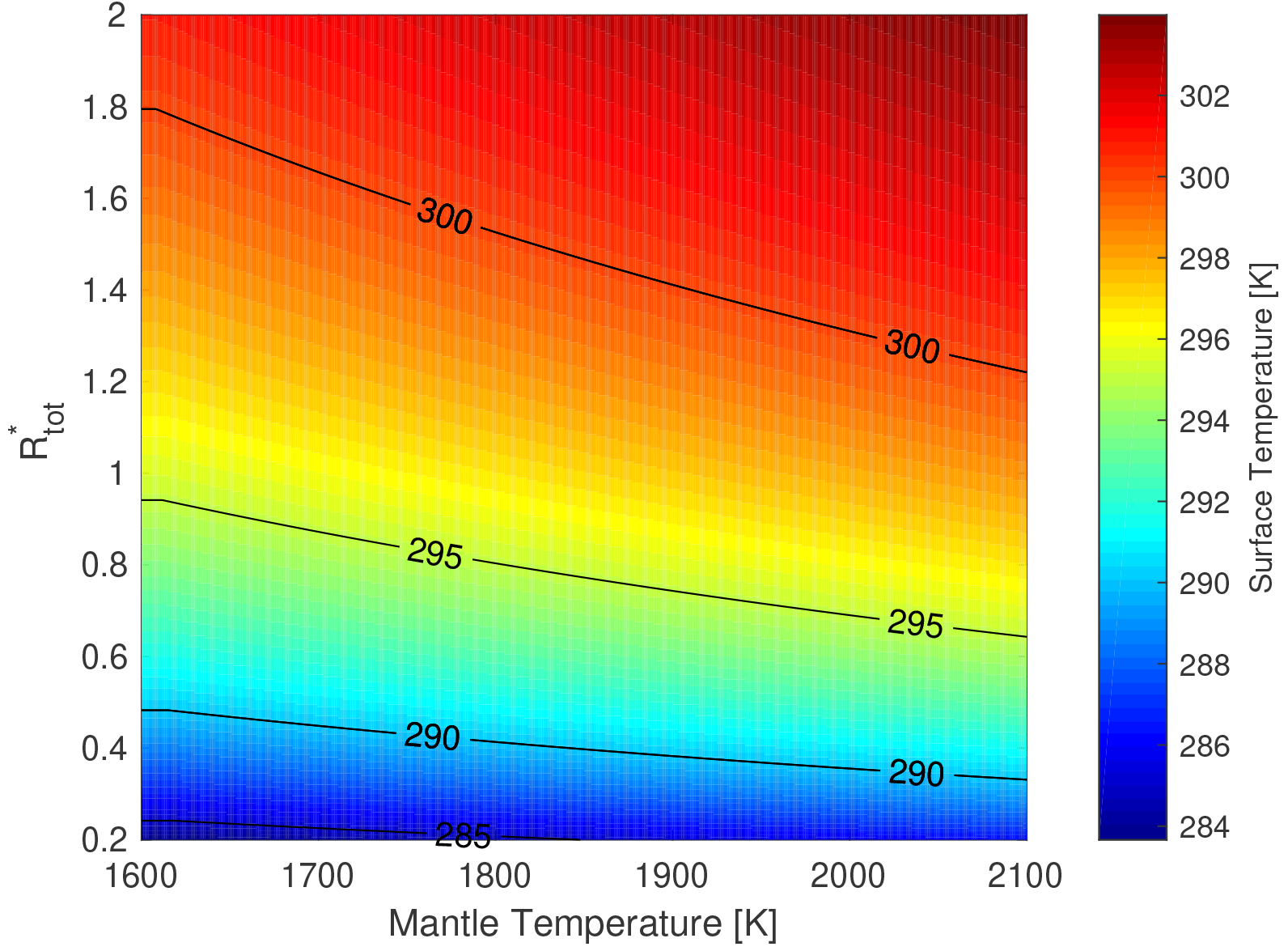}
   \caption{Surface temperature for temperature-dependent weathering as a function of the total CO$_2$ reservoir relative to the Earth $R^*_{tot}$ and the mantle temperature. Although the surface temperature increases with both parameters, temperature-dependent weathering is very efficient in keeping the surface temperature clement.}
   \label{fig_tdep}
\end{figure}

Using a CO$_2$-dependent weathering model (solid and dashed lines in Fig. \ref{fig_ptruns}), the influence of the solar luminosity on the atmospheric CO$_2$ partial pressure is small (compare Figs. \ref{fig_ptruns}a and \ref{fig_ptruns}c). Here, the time-dependence is mainly controlled by the cooling of the mantle. However, the surface temperature is greatly affected by the solar luminosity. While in the early evolution ($<$1 Gyr) the surface temperature decreases with mantle cooling, increasing solar luminosity becomes important after 1 Gyr and the surface temperature increases accordingly.

Using the $T_p$-dependent weathering model (dashed-dotted lines in Fig. \ref{fig_ptruns}), the surface temperature hardly changes with time, and is neither strongly affected by the mantle temperature nor by the solar luminosity. However, the equilibrium CO$_2$ partial pressure strongly decreases as the luminosity increases.

In Fig. \ref{fig_co2dep}, we plot the equilibrium surface temperature as a function of the solar luminosity and the mantle temperature for CO$_2$-dependent weathering using (a) $\alpha=0.23$ and (b) $\alpha=0.25$. The surface temperature increases with both, luminosity and mantle temperature. Therefore, a cooling mantle can partly counteract the increasing luminosity, thereby limiting the net change of the surface temperature. Note that the boundary-layer theory based thermal evolution we use results in a rapid early mantle cooling. However, if thermal evolution models with a slower mantle cooling were applied, mantle cooling would counteract increasing solar luminosity over a longer time span, accordingly.

As discussed above, the solar luminosity hardly affects the steady-state surface temperature if a $T_p$-dependent weathering model is used. However, the equilibrium surface temperature is still a function of the degassing rate, which in turn is sensitive to the total carbon reservoir. In Fig. \ref{fig_tdep}, we plot the equilibrium surface temperature as a function of the total carbon reservoir (relative to the Earth value) and of the mantle temperature for T$_p$-dependent weathering. Note that the model does not account for a limit of the carbon uptake of the crust due to a complete crustal carbonation. However, this limit requires a  total carbon budget of more than $\approx$ two times the Earth's value \citep{Foley:2015}, which is not shown here. As long as this limit is not reached, the surface temperature does not vary significantly for the $T_p$-dependent weathering model.

\subsection{Stagnant lid planets}
\label{sec52}

\begin{figure}
   \centering
   \includegraphics[width=7cm]{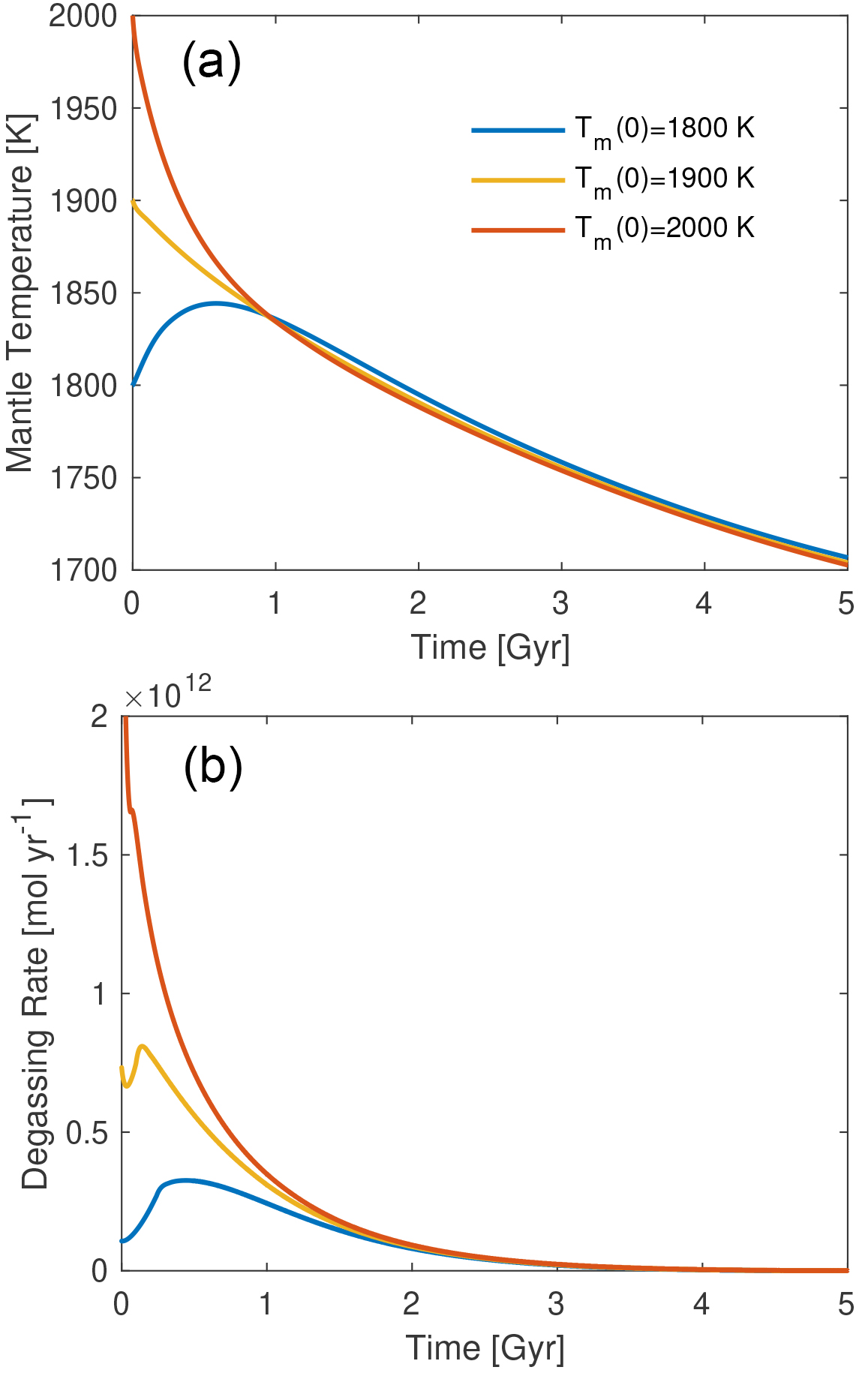}
   \caption{(a) Mantle temperature evolution and (b) degassing rate for a stagnant lid planet with an initial mantle temperature of 1800 K (blue), 1900 K (yellow), and 2000 K (red) using a CO$_2$-dependent weathering model with $\alpha$=0.23.}
\label{fig_slthermal}
\end{figure}

In Fig. \ref{fig_slthermal}, we show thermal evolution models of the mantle of a stagnant lid planet using different initial mantle temperatures of 1800 K (blue), 1900 K (yellow), and 2000 K (red), and the corresponding degassing rates. Neglecting a minor adjustment owed to initial conditions during the first few Myrs, the degassing rate closely follows the mantle temperature evolution. In particular, an initially hot mantle of 1900 K (yellow) or 2000 (red) results in extensive degassing during the first 1 Gyr. In contrast, an initial mantle temperature of 1800 K results in a slightly increase of the mantle temperature and of the degassing rate during the first 0.5 Gyr, followed by a cooling of the mantle and a decreasing degassing rate. After $\sim$ 1 Gyr, the effect of the initial mantle temperature vanishes. In the following, the degassing rate strongly decreases, and between 4 and 5 Gyrs, volcanism ceases completely.

\begin{figure*}
   \centering
   \includegraphics[width=15cm]{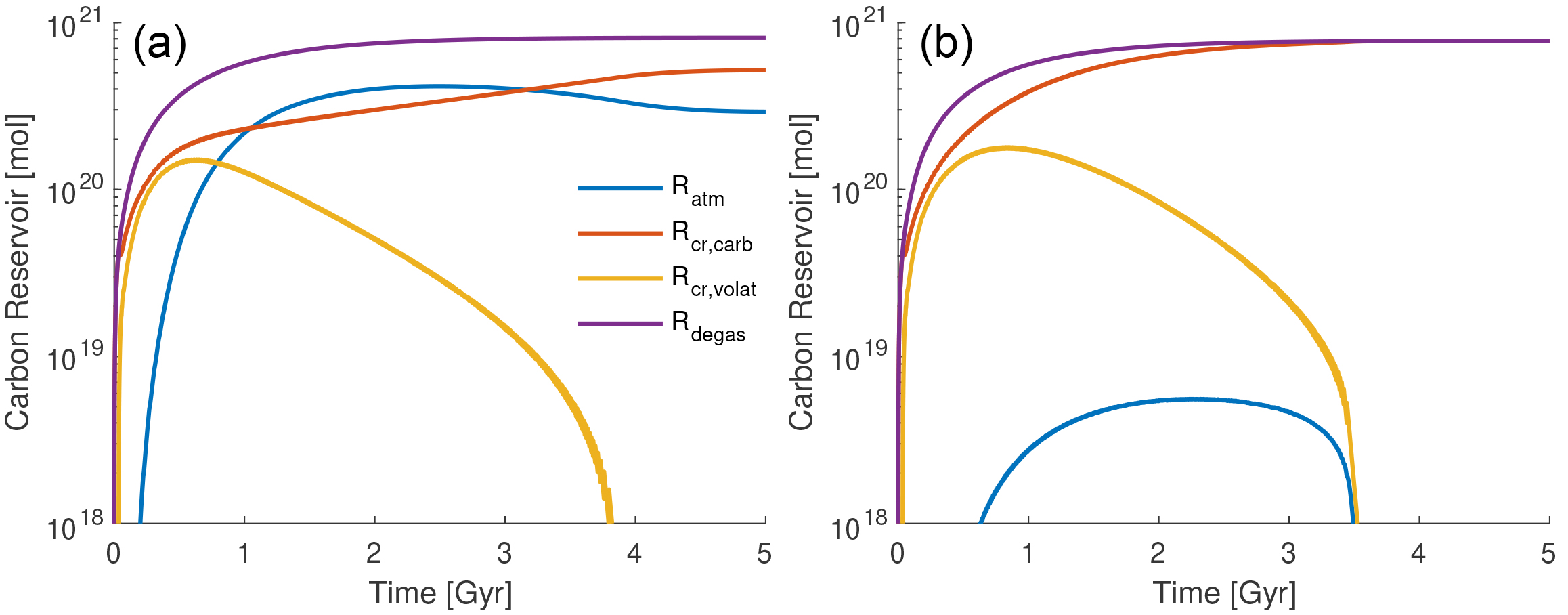}
   \caption{Evolutions of carbon reservoirs for a stagnant lid planet with an initial mantle temperature of 1900 K and (a) CO$_2$-dependent weathering model with $\alpha$=0.23 and (b) $T_p$-dependent weathering model. The accumulated degassed CO$_2$ (magenta) is distributed between the carbonated crust (red), the atmosphere (blue), and the volatile CO$_2$ reservoir in the crustal matrix (yellow).}
\label{fig_slreserv}
\end{figure*}

Modelling the evolution of the crustal and atmospheric CO$_2$ reservoirs of a stagnant lid planet (Fig. \ref{fig_slreserv}), we assume that all the CO$_2$ is initially stored in the mantle. We use an initial mantle temperature of 1900 K and compare (a) a CO$_2$-dependent weathering model with $\alpha$=0.23 with (b) a $T_p$-dependent weathering model. The accumulated degassed CO$_2$ (magenta) is distributed between three reservoirs: the carbonated crust (red), the atmosphere (blue), and the volatile CO$_2$ reservoir in the crustal matrix (yellow), which is fed by decarbonation and gradually releases its CO$_2$ to the atmosphere (implemented to avoid numerical instabilities, see Section \ref{sec4}).

\begin{figure*}
   \centering
   \includegraphics[width=15cm]{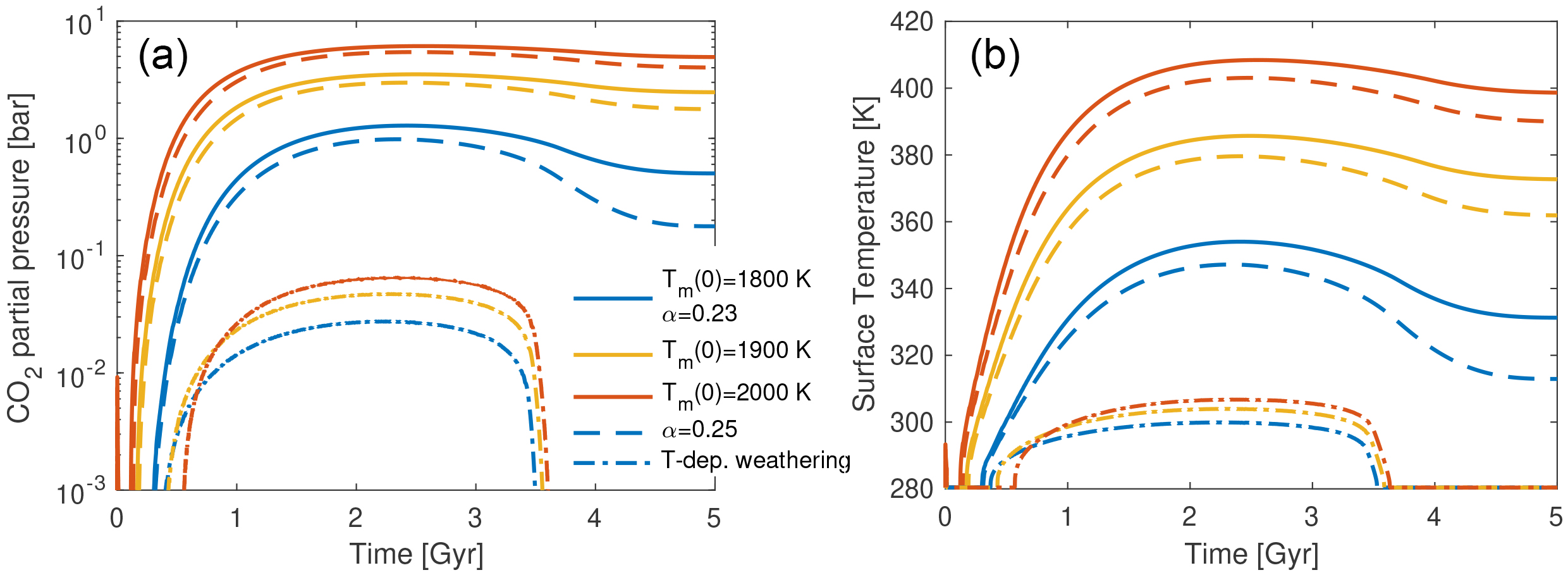}
   \caption{Evolution of (a) atmospheric CO$_2$ and (b) surface temperature for a stagnant lid planet with initial mantle temperatures of 2000 K (red), 1900 K (yellow), and 1800 K (blue). We test CO$_2$-dependent weathering models using $\alpha$=0.23 (solid) and $\alpha$=0.25 (dashed) and a $T_p$ dependent weathering model (dashed-dotted)}
\label{fig_slruns}
\end{figure*}

Using the CO$_2$-dependent weathering model (Fig. \ref{fig_slreserv}a), the degassed CO$_2$ in the early evolution is mainly stored in the crust such that the atmospheric CO$_2$ reservoir remains small. With ongoing degassing and an increase of the crustal carbon reservoir, the CO$_2$ supply rate to the atmosphere increases and with it the atmospheric CO$_2$ reservoir. After 2-3 Gyrs, the atmospheric CO$_2$ reservoir decreases despite an increase of the crustal CO$_2$ reservoir. This is caused by an increase of the decarbonation depth, which results in a reduced decarbonation rate. Since the degassing rate at this time is also small and therefore does hardly contribute to a supply of CO$_2$ to the atmosphere, a net predominance of weathering results in a decrease of the atmospheric CO$_2$. When volcanism ceases completely between 4 and 5 Gyr, the atmospheric CO$_2$ remains approximately constant.

The $T_p$-dependent weathering model (Fig. \ref{fig_slreserv}b) implies an extremely strong weathering feedback and hardly allows any CO$_2$ to remain in the atmosphere. Initial rapid degassing goes along with an increase of the crustal carbon reservoir and the atmospheric CO$_2$ remains small. Note that independent of the weathering scaling, the crustal CO$_2$ reservoir steadily increases as long as partial melting takes place, whereas the atmospheric reservoir decreases in the late evolution as the decarbonation depth increases. This qualitatively differs from stagnant lid models  that do not account for weathering and decarbonation, in which the atmospheric CO$_2$ partial pressure steadily increases (e.g., \citealt{Tosi:2017}).

The effect of different initial mantle temperatures becomes apparent from Fig. \ref{fig_slruns}. We again test initial mantle temperatures of 1800 K (blue), 1900 K (yellow), and 2000 K (red), and CO$_2$-dependent weathering models with $\alpha$=0.23 (solid) and $\alpha$=0.25 (dashed) and a $T_p$-dependent weathering model (dashed-dotted). For the CO$_2$-dependent weathering model, the atmospheric CO$_2$ is greatly affected by the initial mantle temperature. The reason is that any CO$_2$ that is degassed from the mantle cannot be recycled back. A large initial mantle temperature implies rapid CO$_2$ degassing in the early evolution, which in turn causes a rapid increase of the crustal carbon reservoir. Since the crustal reservoir is continuously supplied to the atmosphere via decarbonation, the atmospheric CO$_2$ partial pressure throughout the entire evolution crucially depends on the initial mantle temperature. Since the decarbonation depth depends on the surface temperature, the difference in the atmospheric CO$_2$ partial pressure grows larger. Altogether, differences in the initial mantle temperature of 100 K can cause large differences in the surface temperature at 4.5 Gyr of up to 50 K if a CO$_2$-dependent weathering model is used. In contrast, the $T_p$-dependent weathering model hardly allows any CO$_2$ to remain in the atmosphere, independently of the initial mantle temperature. In the late evolution, the remnant CO$_2$ in the atmosphere is depleted and approaches zero.

\begin{figure*}
   \centering
   \includegraphics[width=18cm]{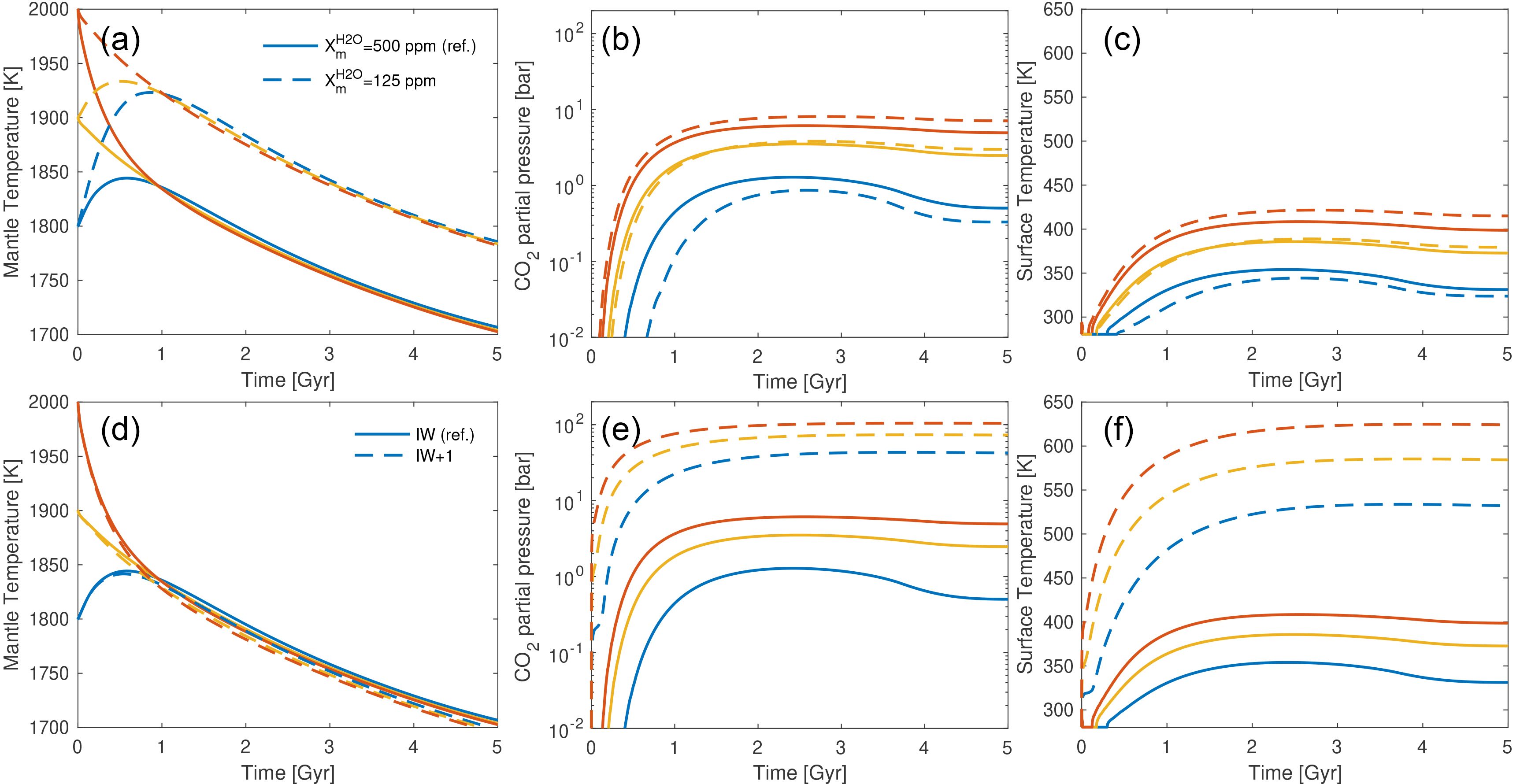}
   \caption{Evolutions of mantle temperature, atmospheric CO$_2$ and surface temperature for a stagnant lid planet with an initial mantle temperature of 1800 K (blue), 1900 K (yellow), and 2000 K (red) using a CO$_2$-dependent weathering model with $\alpha$=0.23. The solid lines refer to the reference model (mantle water concentration of 500 ppm and oxygen fugacity of IW+0). Dashed lines in (a)-(c): Smaller mantle water concentration (125 ppm); Dashed lines in (d)-(e): Larger oxygen fugacity (IW+1).}
\label{fig_slvisc}
\end{figure*}

Particularly for the CO$_2$-dependent weathering model the atmospheric CO$_2$ of a stagnant lid planet is strongly affected by the accumulated mantle degassing, which in turn depends on the mantle rheology and the oxidation state (Appendix \ref{a2}); parameters that are poorly known for extrasolar planets. Fig. \ref{fig_slvisc} illustrates the sensitivity of the model on these parameters using a CO$_2$-dependent weathering model with $\alpha=0.23$ and initial mantle temperatures of 1800 K (blue), 1900 K (yellow), and 2000 K (red). Figs. \ref{fig_slvisc}(a-c) compare the reference case (mantle water concentration of 500 ppm and oxygen fugacity of IW+0, solid) with a case of a reduced mantle water concentration affecting the viscosity (dashed) and Figs. \ref{fig_slvisc}(d-f) compare the reference case (solid) with a model of a larger oxygen fugacity of IW+1 (dashed).

Although the mantle viscosity impacts the evolution of the mantle temperature (Fig. \ref{fig_slvisc}a), an effect on the atmospheric CO$_2$ partial pressure (Fig. \ref{fig_slvisc}b) or surface temperature (Fig. \ref{fig_slvisc}c) is less clear. On the one hand, a large mantle viscosity delays mantle cooling, but on the other hand it results in a more sluggish convection, which reduces the rate of melt production. Whether the combined effect results in a higher or lower atmospheric CO$_2$ partial pressure depends on the initial mantle temperature and is not significant regarding other model uncertainties. In contrast, the oxygen fugacity impacts the evolution of the atmospheric CO$_2$ partial pressure (Fig. \ref{fig_slvisc}e) and the surface temperature (Fig. \ref{fig_slvisc}f) even more strongly than the choice of the initial mantle temperature. This is so because a larger oxygen fugacity enhances the rate at which CO$_2$ is degassed without enhancing the melt production rate.

\section{Discussion}
We presented a model of CO$_2$ recycling for planets with a water-covered surface with and without plate tectonics. In order focus on the main mechanisms and feedbacks, various simplifications have been made.

The complexity of individual subduction zone processes determining their temperature-depth gradients could not be covered within our simple analytical model. In particular, it is difficult to account for parameters that vary from one subduction zone to another, such as the angle and speed of subduction. Instead, we used the present-day distribution of the subduction zone temperature-depth profiles and scaled them according to the evolving mantle temperature. As a result, the fraction of subduction zones avoiding decarbonation increases as the mantle cools. However, this simplification neglects indirect effects of the mantle temperature on subduction zone decarbonation. For example, if the plate speed increases with the mantle heat flux, this may result in limited slab heating in the early evolution and therefore in a smaller decarbonation rate than obtained from our model. Altogether, a more detailed study approaching subduction zone temperatures during the thermal evolution of terrestrial planets is required to obtain a more robust connection between the decarbonation rate and the thermal state of the planet.

Modelling decarbonation, we used the scaling for the decarbonation temperature from \citet{Foley:2018}, based on \citet{Kerrick:2001}, which was derived for dry carbonates. Therefore, we did not account for CO$_2$-release by fluid-mediated reactions \citep{ague:2014}, thereby possibly underestimating the decarbonation flux. In particular, a functional dependence of water release on the temperature-depth profile of a subduction zone, which could then trigger decarbonation, is not captured by our model and may result in a stronger effect of the temperature on the decarbonation rate. Furthermore, we kept the parameter $f$, describing the fraction of subducted CO$_2$ that is not released by arc volcanism, constant. Following Eqs. \ref{eq:ssatm} and \ref{eq:gamma}, the equilibrium CO$_2$ partial pressure directly depends on the stability of carbonates with respect to decarbonation ($\phi$) and arc volcanism ($f$). Future work should consider the effects of temperature and water-release on these parameters in order to improve our understanding of the long-term carbon cycle with mantle cooling.

Although we studied planets with a water-covered surface, we restricted our analysis to planets whose habitability remains controlled by the long-term carbon cycle. For planets with a water content of more than 100 times of that of the Earth and of a water-layer of several hundred km, the long-term carbon cycle becomes unimportant in stabilizing the climate \citep{kite:2018}. The habitability of these planets is rather determined by the partitioning of carbon between the ocean and the atmosphere, by the pressure of the water layer affecting volcanism, and by a possible formation of high-pressure ice phases \citep{noack:2016,kite:2018}.

A quantitative comparison of the results between the plate tectonics model and the stagnant lid model is not straightforward owing to different formulations and parameters of both models. On the one hand, scaling the plate tectonics model to the present-day Earth is advantageous as it reduces the number of free parameters to a minimum (Appendix \ref{a1}). We assumed an Earth-like CO$_2$ degassing rate per melt volume for the present-day mantle carbon reservoir and scaled it with the latter. Using a thermal evolution model then directly allows us to derive conclusions for the early Earth. On the other hand, the stagnant lid requires further assumptions, in particular regarding the rates of crustal production and volcanic degassing (Appendix \ref{a2}), which in turn greatly depends on the oxidation state of the mantle (e.g., \citealt{Hirschmann:2008,Grott:2011,Tosi:2017}). However, despite the differences in the model formulations, the positive feedback connecting surface temperature, decarbonation, and atmospheric CO$_2$ affects the habitability of both model planets. The cooling mantle in the late evolution reduces the atmospheric CO$_2$, regardless of the tectonic mode. For the case of plate tectonics, this is caused by an increasing fraction of subduction zones that avoid decarbonation. For the case of a stagnant lid planet, this is caused by an increasing decarbonation depth.

The dependence of the seafloor weathering rate on CO$_2$ partial pressure and/or temperature used in the model is crucial. In contrast to continental weathering, which is known to be a strong function of the surface temperature \citep{Walker:1981}, a temperature-dependence of seafloor weathering is less certain. Large uncertainties are associated with scaling hydrothermal carbonation with the atmospheric CO$_2$ concentration, as discussed in \citet{Sleep:2001}. The effect of using a different value for $\alpha$ becomes apparent from Eq. \ref{eq:ssatm} since the equilibrium partial pressure of CO$_2$ directly scales with $\alpha$. \citet{Coogan:2015} and \citet{Krissansen-Totton:2017} argued that the water temperature at the bottom of the water layer could control the basalt dissolution rate, thereby implying a strong temperature-dependence of seafloor-weathering. As illustrated in Fig. \ref{fig_ptruns} (for plate tectonics planets) and Fig. \ref{fig_slruns} (for stagnant lid planets), a larger value of $\alpha$ substantially reduces the atmospheric CO$_2$ by increasing the weathering rate, and a temperature-dependent weathering rate stabilizes the surface temperature even more strongly.

\citet{Kadoya:2014} argued that continuous CO$_2$ degassing is a necessary requirement for an Earth-like planet to maintain surface habitability. The reason is that with continuous erosion and weathering on a planet with plate tectonics and emerged land, the carbon sink is always present and continuous degassing is required to keep the balance. This requirement has been applied to stagnant lid planets without continental weathering \citep{Foley:2018}. However, on these planets, the carbon sink is directly coupled to the carbon source. Stopping mantle degassing will inevitably imply a stop of the production of new crust, which will extensively slow down weathering. It remains unclear whether hydrothermal circulation and the seafloor weathering can continue in the absence of volcanic activity. At the latest, a complete carbonation of the upper part of the crust would imply a stop of weathering. If other sinks such as atmospheric escape can be neglected, the change of the atmospheric CO$_2$ partial pressure will be small. In our stagnant lid models (Fig. \ref{fig_slruns}), a decline of the rate of volcanism to a minimum (or even a complete stop) does not imply a rapid decrease of the atmospheric CO$_2$ if a CO$_2$-dependent weathering model is used (solid and dashed lines in Fig. \ref{fig_slruns}). However, the $T_p$-dependent weathering model (dashed-dotted lines in Fig. \ref{fig_slruns}) hardly allows any CO$_2$ to remain in the atmosphere, and therefore, the small atmospheric reservoir can entirely be consumed by weathering when the degassing and decarbonation rates get low (at $\approx$3.5 Gyrs).

Our model is not suited to explore a transition to a potential snowball-state for very small atmospheric CO$_2$ partial pressures. Using a simple climate model following \citet{Walker:1981}, we focus on exploring the ability of planets to maintain a sufficiently low atmospheric CO$_2$ partial pressure in order not to limit their surface habitability by too high temperatures. However, a complete freeze of the surface of an ocean-planet at a low CO$_2$ partial pressure can also limit the surface habitability. \cite{foley:2019} argued that fluctuations in the weathering rate could cause stagnant lid planets with  very small total carbon contents to evolve into a snowball state. He showed that in this case, a continuous exchange of CO$_2$ between the ocean and the atmosphere may be crucial for the planet to recover from this state. Since we found that the initial mantle temperature impacts the atmospheric CO$_2$ of stagnant lid planets in particular if a weak weathering feedback (i.e. a CO$_2$-dependent weathering model) is used, it could also play a role in the ability of a planet to avoid a permanent snowball-state.

The strict separation between the plate tectonics and stagnant lid model throughout the entire evolution of a planet is certainly a simplification. The early Earth may have been in a stagnant lid regime \citep{Debaille:2013}. Changes in the tectonic mode may be caused by changes of the surface temperature \citep{Gillmann:2014}, which would establish a feedback to the climate \citep{Lenardic:2016}. We did not account for temporal changes in the tectonic mode and applied simple parameterized thermal evolution models for both tectonic regimes.

We only considered Earth-sized planets with an Earth-like composition in our model. In particular, the fractions $f$ and $\phi$ that remain stable during subduction are affected by the interior temperature and pressure gradients. If both gradients increase with the planetary mass, the effect of varying the planetary mass on $f$ and $\phi$ will presumably be small. In contrast, if the pressure-dependent viscosity substantially reduces the heat flow for massive Earth-like planets \citep{Stamenkovic:2012}, a large fraction of subducting carbonates will remain stable, thereby reducing the steady-state partial pressure of CO$_2$ in the atmosphere. Furthermore, volcanic degassing may be restricted for massive planets \citep{Kite:2009}. For smaller planets with smaller surface gravitational acceleration such as for example Mars, atmospheric escape may provide an additional important sink to CO$_2$ (e.g., \citealt{Tian:2009}), which is not considered in our model. Certainly, the planetary mass also has an effect on the occurrence of plate tectonics. However, it is not clear whether an increasing planetary mass would make plate tectonics more likely due to a larger shear stress \citep{Valencia:2007} or whether a more buoyant crust would resist subduction \citep{Kite:2009}.

\section{Conclusions}
Since surface erosion on planets with an entirely water-covered surface is negligible, volcanism producing fresh uncarbonated crust is the only mechanism to maintain a carbon sink. Therefore, the weathering rate linearly depends on the rate of volcanism, and is therefore directly coupled to the carbon source. This circumstance becomes particularly important if the stability of carbonated crust with depth is taken into account. Accounting for degassing, seafloor-weathering, decarbonation, and thermal evolution we explored the habitability of planets with a water-covered surface in the plate-tectonics and stagnant lid regime. Our findings are summarized below:

\begin{itemize}

\item[$\bullet$] Coupling seafloor weathering, decarbonation, and interior evolution strengthens the conclusions of previous studies finding that neither emerged land \citep{Abbot:2012,Foley:2015} nor plate tectonics \citep{Tosi:2017,Foley:2018,foley:2019} are required for planets to remain habitable in the long-term. For both tectonic regimes, the CO$_2$ concentration decreases in the late evolution. For plate tectonics planets this is caused by an increasing fraction of subduction zones that avoid crustal decarbonation and for stagnant lid planets this is caused by an increasing decarbonation depth with mantle cooling.

\item[$\bullet$] For plate tectonics planets, the initial mantle temperature is of minor importance. Differences in the initial mantle temperature are negligible after $\approx$1 Gyr in our model, although this time depends on the parameterization of the convective heat transport, however. With continuous carbon recycling into the mantle, the atmosphere-crust-mantle system approaches an equilibrium state, which does not depend on the initial conditions. At the age of the solar system, initial conditions are negligible. However, linking the stability of carbonates in subduction zones to the mantle temperature has implications for the atmospheric CO$_2$ and the surface temperature in the early evolution. Whereas previous studies argue for other greenhouse gases than CO$_2$ in the atmosphere (e.g., \citealt{haqq:2008}) or for a strongly temperature-dependent weathering rate \citep{Krissansen-Totton:2017,krissansen-totton:2018} in order to explain temperate conditions in the early evolution of the Earth, our model indicates that a hotter mantle may partly compensate for the smaller solar luminosity (see Fig. \ref{fig_co2dep}).
  
\item[$\bullet$] For stagnant lid planets, the initial mantle temperature has an effect on the evolution of the atmospheric CO$_2$. This is because a large initial mantle temperature results in a large melt fraction and CO$_2$ degassing rate in the early evolution. Once CO$_2$ has been degassed, it cannot be recycled back into the mantle. Carbonation of the crust can act as a negative feedback to the atmospheric CO$_2$, but since the decarbonation rate depends on the crustal carbon concentration, and because the entire crustal carbon reservoir will be supplied to the atmosphere repeatedly, this negative feedback is weaker than for plate tectonics planets in the long term. Differences in the initial mantle temperature of 100 K can result in differences in the surface temperature at 4.5 Gyr of up to 50 K if a CO$_2$-dependent weathering model is used, thereby greatly impacting the habitability of the planet. In contrast, a temperature-dependent weathering model is very effective in keeping the surface temperature low, independent of the initial mantle temperature. Our results should also hold for stagnant lid planets with an emerged surface: Even if erosion on such a planet takes place, the thereby exposed crust would already have been carbonated at the time it has been produced, such that the weathering rate still linearly depends on the crustal production rate.

\item[$\bullet$] Since the stability of carbonated crust with depth is temperature-dependent, a positive feedback between the surface temperature, decarbonation rate, and atmospheric CO$_2$ is established. Future work studying the difference in the atmospheric CO$_2$ and the runaway greenhouse for planets with different received solar heat fluxes (c.f. \citealt{Driscoll:2013,Foley:2016}) should account for this positive feedback, as it reinforces differences in the atmospheric CO$_2$ partial pressure.
  
\end{itemize}
   
\begin{acknowledgements}
      We thank Brad Foley for his constructive review, which greatly helped to improve this paper. DH has been supported through the NWA StartImpuls and NT has been supported by the Helmholtz Association (project VH$-$NG$-$1017) and by the German research foundation (DFG) through the priority programs 1922 “Exploring the diversity of extrasolar planets” (TO 704/3-1) and 1883 “Building a habitable Earth” (TO 704/2-1).
\end{acknowledgements}

\begin{appendix}

\section{Thermal evolution model for plate tectonics planets}
\label{a1}

Calculating the thermal evolution of a plate tectonics planet, we follow boundary layer theory and a parameterization of the heat transport by convection \citep{Schubert:2001,Davies:2007,Honing:2016}. The evolution of the mantle temperature is calculated by integrating the energy conservation equation

\begin{linenomath}
\begin{equation}
    \rho_m c_m V_m  \frac{\mathrm{d}T_m}{\mathrm{d}t}=
     -q_u(t)A_m
     +q_l(t)A_c
     +Q_m(t)V_m,
    \label{eqa11}
\end{equation}
\end{linenomath}
where $\rho_m$ is the mantle density, $c_m$ the heat capacity, $V_m$ the volume, $A_m$ the surface area, and $A_c$ the core surface area. The mantle heat production rate is
\begin{linenomath}
\begin{equation}
    Q_m(t)=Q_0\sum_i Q_i'\exp{\left(-\frac{log(2)}{h_i}(t-t_E)\right)},
    \label{eqa12}
\end{equation}
\end{linenomath}
where $t$ represents the time in Gyr, $t_E$=4.5 Gyr is the present-day age of the Earth, $Q_0$ is the total present-day heat production rate, $Q'_1...Q'_4$ are the relative present-day heat production rates and $h_1...h_4$ the halt-life times of $^{238}$U, $^{235}$U, $^{232}$Th, and $^{40}$K as given in \citet{Korenaga:2008}.

The heat fluxes $q_u$ and $q_l$ through the upper and lower thermal boundary layers are given by
\begin{linenomath}
\begin{equation}
    q_u=k\frac{T_m-T_s}{\delta_u}
    \label{eqa13}
\end{equation}
\end{linenomath}
and

\begin{linenomath}
\begin{equation}
    q_l=k\frac{T_c-T_b}{\delta_l},
    \label{eqa14}
\end{equation}
\end{linenomath}
where $k$ is the thermal conductivity, $\delta_u$ and $\delta_l$ are the upper and lower boundary layer thicknesses, and $T_s$, $T_c$ and $T_b$ are the surface, core and bottom mantle temperatures. We use a local instability criterion following \citet{Stevenson:1983} and set the Rayleigh number $Ra$ at the upper boundary layer equal to the critical Rayleigh number $Ra_{crit}$ to calculate the thicknesses of the boundary layer
\begin{linenomath}
\begin{equation}
    Ra_u=\frac{\overline{\alpha}\rho_m^2 c_p g(T_m-T_s)\delta_u^3}{k\eta}=Ra_{crit},
    \label{eqa15}
\end{equation}
\end{linenomath}
where $g$ is the gravitational acceleration, $\eta$ the viscosity, $\overline\alpha$ the thermal expansion coefficient, and $c_p$ the mantle heat capacity. For simplicity, we assume a symmetric temperature profile throughout the mantle and set the thickness of the lower thermal boundary layer equal to the thickness of the upper boundary layer, i.e., $\delta_l=\delta_u$. The viscosity at the upper thermal boundary layer is temperature-dependent:
\begin{linenomath}
\begin{equation}
     \eta(T_m)=\eta_E\exp{\left(\frac{E^*}{R}\left(\frac{1}{T_m} - \frac{1}{T_{m,E}}\right)\right)},
    \label{eqa16}
\end{equation}
\end{linenomath}
where $\eta_E$ and $T_{m,E}$ are the present-day Earth mantle viscosity and mantle temperature, $E^*$ the activation energy, and $R$ the gas constant. The adiabatic temperature profile in the mantle is calculated using the linearized form

\begin{linenomath}
\begin{equation}
    T_b=T_m\left(1+\frac{\overline\alpha g d_m}{c_p}\right),
    \label{eqa17}
\end{equation}
\end{linenomath}
where $d_m$ is the mantle thickness. The evolution of the core temperature $T_c$ is calculated by

\begin{linenomath}
\begin{equation}
    \frac{\mathrm{d}T_c}{\mathrm{d}t}=-\frac{q_b A_c}{\rho_c c_c V_c},
    \label{eqa18}
\end{equation}
\end{linenomath}
where $A_c$, $V_c$, $\rho_c$, and $c_c$ are the surface area, volume, density and specific heat capacity of the core. Parameter values are given in Tab. \ref{tab2}.

\begin{table}
\caption[]{Parameter values for the thermal evolution models.}
\label{tab2}
\centering                          
\begin{tabular}{l p{3.6cm}l l}
\hline
Parameter & Description & Value \\
\hline
$\rho_m$       & Mantle density   & 3.5 $\cdot 10^{3}$ kg m$^{-3}$   \\
$c_m$      & Mantle heat capacity  & 1.1 $\cdot 10^{3}$ J K$^{-1}$ kg$^{-1}$ \\
$\rho_c c_c$   & Core density \newline times heat capacity  & 3.6 $\cdot 10^6$ J K$^{-1}$  m$^{-3} $  \\
$\rho_{cr} c_{cr}$  & Crust density \newline times heat capacity  & 3.2 $\cdot 10^{6}$ J K$^{-1}$  m$^{-3}$\\
$k$  & Thermal conductivity & 4 W m$^{-1}$ K$^{-1}$ \\
$\overline{\alpha}$  & Thermal expansion  & 2 $\cdot 10^{-5}$  K$^{-1}$  \\
$g$   & Gravitational acc.   & 9.81 m s$^{-2}$   \\
$Ra_{crit}$  & Crit. Rayleigh Number   & 450        \\
$V_c$         & Core volume   & 1.7 $\cdot 10^{20}$  m$^{3}$      \\
$V_m$       & Mantle volume     & 9.1 $\cdot 10^{20}$ m$^{3}$      \\
$A_c$    & Core surface area  & 1.5 $\cdot 10^{14}$  m$^2$   \\
$A_m$    & Mantle surface area  & 5.1 $\cdot 10^{14}$  m$^2$   \\
$E^*$ & Activation energy  & 3 $\cdot 10^{5}$  J mol$^{-1}$  \\
$V^*$ & Activation volume  & 4  $\cdot 10^{-6}$ m$^3$ mol$^{-1}$  \\
$\eta_E$ & Present-day Earth mantle viscosity (Eq. \ref{eqa16}) & $10^{21}$ Pa s  \\
$A^{'}$ & Viscosity pre-factor (Eq. \ref{eqa24}) & 6.127  $\cdot 10^{10}$ Pa s  \\
$R$     & Gas constant      & 8.314    \\
$T_{c,ini}$  & Initial core temperature    & 3500 K    \\
$f_p$   & Surface fraction of hot upwellings with melting & 0.01   \\
$R$  & Gas constant     & 8.314             \\
$u_0$   & Charact. velocity scale      & 2 $\cdot 10^{-12}$ m s$^{-1} $  \\
$L$  & Latent heat of melting  & 5 $\cdot 10^5 $ J kg$^{-1}  $    \\
\hline
\end{tabular}
\end{table}

\section{Thermal evolution model for stagnant lid planets}
\label{a2}
We use a parameterized model of mantle convection in the stagnant lid regime and CO$_2$ degassing that has been explained in detail in \citet{Tosi:2017}. Therefore, we only give a brief overview of the model here. The conservation of energy is given by

\begin{linenomath}
\begin{equation}
\begin{split}
    \rho_m c_m V_l & (1+{St})  \frac{\mathrm{d}T_m}{\mathrm{d}t} = \\\
    &-\left(q_l+\left(\rho_{cr} L+\rho_{cr} c_{cr} (T_m-T_l)\right) \frac{\mathrm{d}D_{cr}}{\mathrm{d}t}\right) A_l \\\
  & +q_c A_c+Q_m V_l
     ,
\end{split}
\label{eq:tosi2}
\end{equation}
\end{linenomath}
where ${St}$ is the Stefan number controlling the release and consumption of latent heat upon melting and solidification, $q_l$, $A_l$, and $V_l$ are the heat flux through the lid and its area and volume, $\rho_{cr}$, $c_{cr}$, and $D_{cr}$ are the crustal density, heat capacity, and thickness, and $L$ is the latent heat of melting. The evolution of the stagnant lid is given by
\begin{linenomath}
\begin{equation}
\begin{split}
    \rho_m c_m & (T_m-T_l) \frac{\mathrm{d}D_l}{\mathrm{d}t}= \\\ & -q_l+\left(\rho_{cr} L+\rho_{cr} c_{cr} (T_m-T_l )\right)  \frac{\mathrm{d}D_{cr}}{\mathrm{d}t} -k \left.\frac{\partial T}{\partial r}\right\rvert_{r=R_l},
\end{split}
    \label{eqa22}
\end{equation}
\end{linenomath}
where $D_l$ is the lid thickness and $\frac{\partial T}{\partial r}$ is the radial temperature gradient at the base of the lid (i.e., at the radius $r=R_l$), which is calculated using the steady-state heat conduction equation
\begin{linenomath}
\begin{equation}
    \frac{1}{r}  \frac{\partial}{\partial r} \left(r^2 k \frac{\partial T}{\partial r}\right)+Q_l=0,
    \label{eq:tosi4}
\end{equation}
\end{linenomath}
where $r$ is the radial coordinate, and $Q_l$ and $k$ are the heat production rate and the thermal conductivity of the lid.
The heat fluxes $q_u$ and $q_l$ are obtained following boundary layer theory (Appendix \ref{a1}). We account for the water-dependence of the mantle viscosity as explained in \citet{Tosi:2017}
\begin{linenomath}
\begin{equation}
    \eta=\frac{A'}{X_m^{H_2 O}}\exp{\left(\frac{E^*+p_m V^*}{RT_m}\right)},
    \label{eqa24}
\end{equation}
\end{linenomath}
where $A'$ is the pre-exponential factor and $X_m^{H_2 O}$ the water concentration in ppm (for the reference models, we use $X_m^{H_2 O}$=500 ppm), $E^*$ is the activation energy, $V^*$ is the activation volume, and $p_m$ is the lithostatic pressure at the upper mantle. We follow \citet{Grasset:1998} and calculate the lid temperature by

\begin{linenomath}
\begin{equation}
    T_l=T_m-\Theta\frac{RT_m^2}{E^*},
    \label{eqa25}
\end{equation}
\end{linenomath}
where $\Theta=2.9$ accounts for spherical geometry.

Calculating the local melt fraction $\Phi(r)$, we compare the mantle temperature profile $T(r)$ with the solidus $T_{sol}(r)$ and liquidus $T_{liq}(r)$ temperatures as given in \citet{Katz:2003}:

\begin{linenomath}
\begin{equation}
    \Phi(r)=\frac{T(r)-T_{sol}(r)}{T_{liq}(r)-T_{sol}(r)}
    \label{eq:tosi15}
\end{equation}
\end{linenomath}
The volume-averaged melt fraction is calculated as

\begin{linenomath}
\begin{equation}
    \overline\Phi=\frac{1}{V_{\Phi}} \int_{V_\Phi} \Phi(r)d\Phi,
    \label{eq:tosi16}
\end{equation}
\end{linenomath}
where $V_\Phi$ is the melt volume. The crustal growth rate is calculated as

\begin{linenomath}
\begin{equation}
    \frac{\mathrm{d}D_{cr}}{\mathrm{d}t}=f_p u\overline\Phi \frac{V_\Phi}{4\pi R_p^3},
    \label{eq:tosi17}
\end{equation}
\end{linenomath}
where $R_p$ is the planet radius, $f_p$ is a constant describing the fraction of the surface covered by hot plumes in which partial melting takes place, and $u$ is the convective velocity

\begin{linenomath}
\begin{equation}
    u=u_0 \left(\frac{Ra}{Ra_{crit}}\right)^{2\beta},
    \label{eqa29}
\end{equation}
\end{linenomath}
where $u_0$ is the characteristic mantle velocity scale and $\beta=1/3$ is a constant. If the crustal thickness exceeds the lid thickness, we set $D_{cr}=D_{l}$, causing crustal recycling into the mantle.

We account for partitioning of incompatible elements between the crust and the mantle. The concentration of a trace element in the liquid phase $X_{liq}$ relative to its bulk mantle concentration $X_{m}$ is calculated from

\begin{linenomath}
\begin{equation}
    \frac{X_{liq}}{X_{m}}=\frac{1}{\Phi} \left(1-(1-\Phi)^{1/\delta}\right),
    \label{eq:tosi19}
\end{equation}
\end{linenomath}
where $\delta=10^{-3}$ is the partition coefficient for heat producing elements \citep{blundy:2003}. Combining Eqs. \ref{eq:tosi15} with \ref{eq:tosi19} yields the average concentration of incompatible elements in the crust, and the total rate at which incompatible elements are extracted is set proportional to the crustal production rate (Eq. \ref{eq:tosi17}). The volumetric heating rate of the mantle as needed in Eq. \ref{eq:tosi2} is then calculated dependent on the respective abundance of heat producing elements in the mantle, and the heat production in the lid (Eq. \ref{eq:tosi4}) is derived from the heating rates in the crust and mantle using their relative volume fractions. A more thorough explanation of the thermal evolution model is given in \citet{Tosi:2017}.

To calculate the CO$_2$ degassing rate, we follow the redox melting and degassing model of \citet{Grott:2011} and assume that the mantle oxygen fugacity is sufficiently low for carbon to be available in its reduced form. Upon partial melting, some graphite is dissolved in the melt and subsequently degassed. The concentration of CO$_2$ in the melt depends on the carbonate concentration, which in turn is a function of the oxygen fugacity as given in \citet{Grott:2011} and \citet{Tosi:2017}, using scaling constants appropriate for Hawaiian basalts \citep{holloway:1998}. We keep the oxygen fugacity fixed at the iron-w\"ustite buffer IW but also test an oxygen fugacity of IW+1. The CO$_2$ degassing rate directly follows from the concentration of CO$_2$ extracted from the mantle and the melt fraction as given in Eq. \ref{eq:tosi16}, using a fixed extrusive-to-intrusive volcanism ratio of 2.5 \citep{Grott:2011}. More details on the CO$_2$ degassing model used for stagnant lid planets can be found in \citet{Tosi:2017}.


\section{Climate model}
\label{a3}

\begin{table}
\caption[]{Parameter values for the climate model.}
\label{tab3}
\centering                          
\begin{tabular}{l p{3.2cm}l l}        
\hline
Parameter & Description & Value \\
\hline
$T_{s,E}$ & Present-day Earth \newline surface temperature & 285 K    \\
$S_E$     & Solar constant                        & 1360 Wm$^{-2}$  \\      
$t_E$     & Age solar system               & 4.5 Gyr    \\
$A$       & Albedo                                & 0.31    \\
$\sigma$ & Stefan-Boltzmann constant     & 5.67$\cdot 10^{-8}$  W m$^{-2}$ K$^{-4}$ \\
\hline
\end{tabular}
\end{table}

To test the feedback cycle between mantle temperature, atmospheric CO$_2$, and surface temperature, we apply a simple climate model following \citet{Walker:1981}. We relate the atmospheric CO$_2$ partial pressure to the surface temperature as

\begin{linenomath}
\begin{equation}
    T_s=T_{s,E}+2\Delta T_{eff}+4.6\left(\left(\frac{P_{CO_2}}{P_{CO_2,E}}\right)^{0.346}-1\right),
    \label{eqa31}
\end{equation}
\end{linenomath}
where $T_{s,E}$ is the present-day Earth surface temperature, $P_{CO_2}$ the atmospheric CO$_2$ partial pressure (and $P_{CO_2,E}$ the present-day Earth value), and the difference of the effective temperature is given by

\begin{linenomath}
\begin{equation}
    \Delta T_{eff}=\left(\frac{S(1-A)}{4\sigma}\right) ^{1/4}-\left(\frac{S_E (1-A)}{4\sigma}\right)^{1/4},
    \label{eqa32}
\end{equation}
\end{linenomath}
where $\sigma$ is the Stefan-Boltzmann constant, $A$ the albedo, and the solar flux following \citet{Gough:1981} is given by:

\begin{linenomath}
\begin{equation}
    S=S_E \left(1+\frac{2}{5} \left(1-\frac{t}{t_E}\right)\right)^{-1},
    \label{eqa33}
\end{equation}
\end{linenomath}
where $t$ is the time in Gyr ($t_E$=4.5 Gyr is the age of the present-day solar system) and $S_E$ the present-day Earth solar flux. Parameter values are given in Tab. \ref{tab3}.

\end{appendix}

\bibliographystyle{aa} 
\bibliography{tba}

\end{document}